\documentclass[%
reprint,
superscriptaddress,
amsmath,amssymb,
prl,
longbibliography,
nofootinbib
]{revtex4-2}

\usepackage{graphicx}% Include figure files
\usepackage{dcolumn}% Align table columns on decimal point
\usepackage{bm}% bold math
\usepackage{natbib}
\usepackage{braket}
\usepackage{xcolor}
\usepackage{nicefrac}
\usepackage{array}
\usepackage{caption}
\usepackage{subcaption}
\usepackage[T1]{fontenc}
\usepackage{tikz}

\graphicspath{{./images/}}
\newcommand{\abbv}[2]{\item \textbf{#1: } #2}

\newcommand{\tcirc}[1]{\textcolor{blue}{\raisebox{.5pt}{\textcircled{\raisebox{-.9pt} {#1}}}}}

\newcommand{\mics}{\,$\mu$s\:}

\newcommand{\change}[1]{{#1}}

\newcommand\copyrighttext{%
  \footnotesize \textcopyright 2022 IEEE. Personal use of this material is permitted.
  Permission from IEEE must be obtained for all other uses, in any current or future
  media, including reprinting/republishing this material for advertising or promotional
  purposes, creating new collective works, for resale or redistribution to servers or
  lists, or reuse of any copyrighted component of this work in other works.
  }
\newcommand\copyrightnotice{%
\begin{tikzpicture}[remember picture,overlay]
\node[anchor=south,yshift=10pt] at (current page.south) {\fbox{\parbox{\dimexpr\textwidth-\fboxsep-\fboxrule\relax}{\copyrighttext}}};
\end{tikzpicture}%
}
\begin{document}
% \history{Date of publication xxxx 00, 0000, date of current version xxxx 00, 0000.}
%\doi{10.1109/TQE.2020.DOI}

%\title{Design and analysis of digital communication within an MPSoC-based control system for trapped-ion quantum computing}
\title{Design and analysis of digital communication within an SoC-based control system for trapped-ion quantum computing}
\author{Nafis Irtija}
\email{nafis@unm.edu}
\affiliation{University of New Mexico, Albuquerque, NM, USA}
\author{Jim Plusquellic}
\affiliation{University of New Mexico, Albuquerque, NM, USA}
\author{Eirini Eleni Tsiropoulou}
\affiliation{University of New Mexico, Albuquerque, NM, USA}
\author{Joshua Goldberg}
\affiliation{Sandia National Laboratories, Albuquerque, NM, USA}
\author{Daniel Lobser}
\affiliation{Sandia National Laboratories, Albuquerque, NM, USA}
\author{Daniel Stick}
\affiliation{Sandia National Laboratories, Albuquerque, NM, USA}

\date{\today}
% \tfootnote{This work was funded by Q-SEnSE: Quantum Systems through Entangled Science and Engineering (NSF QLCI Award OMA-2016244) and by QSA: Quantum Systems Accelerator (U.S. Department of Energy, Office of Science, National Quantum Information Science Research Centers).}
% %\tfootnote{This paragraph of the first footnote will contain support information, including sponsor and financial support acknowledgment. For example, ``This work was supported in part by the U.S. Department of Commerce under Grant BS123456.''}

% \markboth
% %{Nafis Irtija \headeretal: Preparation of Papers for IEEE Transactions on Quantum Engineering}
% {Nafis Irtija \headeretal: Design and analysis of digital communication within an MPSoC-based control system for TIQC}
% {Nafis Irtija \headeretal: Design and analysis of digital communication within an MPSoC-based control system for TIQC}

% \corresp{Corresponding author: Nafis Irtija (email: nafis@unm.edu).}

\begin{abstract}

Electronic control systems used for quantum computing have become increasingly complex as multiple qubit technologies employ larger numbers of qubits with higher fidelity targets. Whereas the control systems for different technologies share some similarities, parameters like pulse duration, throughput, real-time feedback, and latency requirements vary widely depending on the qubit type. In this paper, we evaluate the performance of modern System-on-Chip (SoC) architectures in meeting the control demands associated with performing quantum gates on trapped-ion qubits, particularly focusing on communication within the SoC. A principal focus of this paper is the data transfer latency and throughput of several high-speed on-chip mechanisms on Xilinx multi-processor SoCs, including those that utilize direct memory access (DMA).  They are measured and evaluated to determine an upper bound on the time required to reconfigure a gate parameter. Worst-case and average-case bandwidth requirements for a custom gate sequencer core are compared with the experimental results. The lowest-variability, highest-throughput data-transfer mechanism is DMA between the real-time processing unit (RPU) and the PL, where bandwidths up to 19.2 GB/s are possible. For context, this enables reconfiguration of qubit gates in less than 2\mics\!, comparable to the fastest gate time.  Though this paper focuses on trapped-ion control systems, the gate abstraction scheme and measured communication rates are applicable to a broad range of quantum computing technologies. 

\end{abstract}

% \begin{keywords}
% trapped-ion, qubits, quantum computing, SoC-based FPGA control system
% \end{keywords}

% \titlepgskip=-15pt

\maketitle
\copyrightnotice

\section{Introduction}
\label{sec:introduction}

The most common way of performing quantum gates \cite{nielsen:2010} in a trapped-ion quantum computer (TIQC) uses modulated laser pulses that interact with the atomic energy levels of the ions.  These pulses are normally generated with radiofrequency (RF) signals that modulate the frequency, phase, and amplitude of light using acousto-optic modulators (AOMs) \cite{clark:2021, pogorelov:2021}.  Typical hardware components for generating these RF signals include arbitrary waveform generators (AWGs), direct-digital synthesizer (DDS) modules, and field programmable gate arrays (FPGA) that directly drive digital-to-analog converters (DACs) using an AWG-type architecture, soft-core DDSs, or some combination of the two. Other interaction mechanisms that involve RF or microwave signals delivered to ions via antennae or electrodes incorporated directly into the ion traps can also be used to perform gates, and use similar control electronics as those needed for AOMs.

In this paper we describe the digital side of a control system for generating RF signals that drive quantum gates in a TIQC \cite{clark:2021}, focusing on data bandwidth requirements for supporting RF operating frequencies, synchronization, phase coherence, and feedback-based calibration. This latter need for near real-time feedback provides a context to calculate communication requirements within the control system.  Our goal is to arbitrarily reconfigure gate parameters based on preceding measurements within a time period that is on the order of the fastest gate time (assumed here to be a 1\mics single qubit gate).
Given a particular parameterization scheme, this establishes a target communication rate and latency for different parts within the SoC, i.e. the integrated multi-core microprocessors with high-speed on-chip data buses and memory mechanisms shared with the programmable fabric. We categorize control operations into those requiring fixed-cycle-count time intervals, those that are deadline-based, and those that have soft real-time constraints for mapping into the programmable logic (PL) and processing system (PS) components of the SoC.
This requirement may seem overly stringent, but it has the advantage that it could prevent correlated errors that are correctable by control hardware from jeopardizing error correcting circuits.  An example would be tuning the amplitudes of RF signals that are driving AOMs to correct for power changes in a laser that supports multiple qubits.

Our co-design approach expresses complex hardware-centric features using software-based constructs, for example
the data and control signals are constructed by an application running on a processor(s) and are transferred to the control system in the PL using direct memory access (DMA) by a real-time processor. 
Beyond the benefit of abstracting away low-level operational details, this approach also reduces bandwidth requirements across the software-hardware interface \cite{fu:2019, cross:2021}. The objective of our hardware performance evaluation is to explore a variety of specialized hardware acceleration features that can be leveraged to further improve performance across the TIQC hardware-software interface. This paper does not analyze analog errors associated with the control system, which have an important effect on gate fidelity and must be addressed when considering the entire control system \cite{bardin:2021,vandijk:2019}. 

\begin{figure}
   \centering
   \includegraphics[width=\columnwidth]{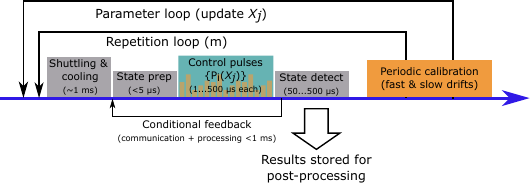}
   \caption{A typical experimental cycle for a trapped-ion quantum computer.  A gate sequence consisting of preparation, a series of $n$ control pulses (where $i \in \{1,n\}$), and measurement/state detection, which is repeated $m$ times for statistics.  There are $o$ different parameters that define these individual pulses (where $j \in \{1,o\}$), which are updated in the parameter loop.  Periodic calibration occurs on fast and slow time scales to correct for drift and other hardware errors. The electronics described here are also capable of conditional feedback such that pulse parameters can be modified based on individual or groups of measurements.  More commonly the results are stored for post-processing.}
   \label{fig:expcycle}
\end{figure}

A block diagram illustrating high level experimental operations with multiple control loops is shown in Fig. \ref{fig:expcycle}. The loops are required to enable sufficient statistics to be gathered as well as tune gate sequences in subsequent iterations by feeding back on results obtained from past iterations. All of these loops operate 
on a deadline-based schedule with an upper-bound on time constraints,
and the effectiveness of the feedback-driven tuning operations is ultimately limited by the bandwidth of the hardware-software interface. 

Our design is meant to accommodate a wide array of tasks needed for typical day-to-day operation of a TIQC, including the most challenging scenario of running an algorithm that requires many consecutive operations.
In contrast to classical high performance computing in which checkpoints can be used to store intermediate calculations to recover from unexpected failures \cite{liu:2008}, quantum computers cannot classically store intermediate states.  In addition, simply to preserve those quantum states requires continuous quantum error correction (QEC) \cite{chamberland:2017} and calibration. Control errors that occur outside of the electronic control system, like fluctuations in laser intensity or external magnetic fields can be corrected by the control system as long as they are detected, the correction is calculated, and the control parameters for the many affected qubits are updated, all in less time than it takes for the next gate to be applied within a round of syndrome extraction. Therefore, in this research we have focused considerable effort on measuring the communication limitations that would bound the reconfiguration time of the control system.  There are other approaches to dealing with this scenario, like restarting the algorithm, dynamical decoupling, or distributing control lines 
in such a way as to not risk highly-correlated errors within a logical qubit, but these may come at considerable cost and would ideally be made unnecessary by a sufficiently fast control system.

With typical gate times ranging from 1\mics to 1 ms, the timing requirements for a TIQC control system are significantly different than other technologies that have gates that can be 100 times faster. This relaxes some performance requirements; for instance the rotation angle applied during a single qubit gate is proportional to the interaction time of the pulse, so a 1\% angular rotation error for a 1\mics gate can be achieved if there is less than 10 ns error in the gate duration, posing less stringent absolute timing demands than technologies with faster gates. The conditional feedback described above is another example in which the slower trapped-ion system relaxes control system requirements.

The Xilinx Zynq UltraScale+ radiofrequency SoC (RFSoC) device, embedded within a Xilinx ZCU111 evaluation board \cite{xilinxZCU}, is used as the experimental platform in our evaluation. The digital components integrated onto this device include multiple CPUs, shared memory, and a PL fabric. The PS and PL each connect to a dedicated 4 GB bank of DDR4 (DRAM) memory. The PS side includes an Arm Cortex-A53 64-bit quad-core application processing unit (APU) and a Cortex-R5 32-bit dual-core RPU, local caches and on-chip scratch memory, all interconnected with a complex Arm Advanced eXtensible Interface (AXI) switch network to enable interprocessor communication and high-speed communication channels between the PS and PL sides. The ZCU111 is a mixed-signal device, integrating dedicated, high-speed analog RF components, in particular, eight 4 GSPS 12-bit RF analog-to-digital converters (ADCs) and eight 6.5 GSPS 14-bit RF digital-to-analog converters (DACs). The applications that have driven the marketing and development of the ZCU111 include 5G Wireless, Next-Generation ADAS, and Industrial Internet-of-Things,
but many features of the architecture are well-suited for quantum computing as well.

We also compare the performance in some cases with a less expensive ZCU102 MPSoC evaluation board that possesses a nearly identical digital processing architecture to the ZCU111. In particular, the latency and throughput characteristics of both devices are presented for the DMA transfer mechanisms investigated in this paper, as an illustration of the performance benefit that is attainable when using a faster DDR, which is a feature of the ZCU111. We envision a larger, multi-SoC qubit system that can utilize the ZCU102 for system-level coordination and synchronization among a set of ZCU111s. Therefore, the performance characteristics of the ZCU102 are also relevant for quantum computing systems. 

Hardware design and analysis covered in this paper includes:
\begin{itemize}
    \item The hardware and software control elements of a custom DDS for a TIQC system are described, along with an analysis of the worst case and average case throughput that are required between the PS and PL sides of the ZCU111.
    \item An analysis of the throughput and latency associated with a set of four distinct communication mechanisms within the ZCU boards is presented, as well as an analysis of the variability associated with these channels.
    \item A feasible mapping of TIQC communication channel requirements and those available within the digital architecture of the ZCU boards is presented and the tradeoffs and limitations discussed.    
\end{itemize}

The remainder of this paper presents related work (Section \ref{sec:background}), an overview of the overall TIQC system architecture and task partitioning (Section \ref{sec:architecture}), a detailed description of the characteristics, functionality and requirements of gate sequence generation implemented by custom PL components of the control system (Section \ref{sec:dds}), a description of the experimental setup and an analysis of throughput and latency associated with four high-speed, on-chip communication mechanisms within the ZCU111 (Section \ref{sec:com}), and a presentation of a feasible mapping strategy between ZCU111 communication mechanisms and the required communication channels within the TIQC system (Section \ref{sec:limits}). 
\section{Related Work}
\label{sec:background}

Quantum computing experiments that use both custom \cite{mount:2015, keitch:2017} and commercial \cite{schafer:2018, kasprowicz:2020} FPGA-based control systems have been demonstrated over the last decade. FPGA-based architectures for quantum communication have also been proposed \cite{stanco:2022}.  Whereas earlier experiments emphasized the flexibility of generating control pulses, more recent hardware has focused on scaling and its concomitant challenges.  For example, \cite{messaoudi:2020} describes a modular system that uses PXIe modules for arbitrary waveform generation and ADC sampling that can support extending the number of controllable qubits while maintaining nanosecond level synchronization. 
Another example is the Virtex-7 FPGA custom platform proposed in \cite{qin:2019} for control of spin-based qubits that includes a 1 GS/s AWG, an 8-channel pulse/sequence generator, a 2-channel ADC and a 2-channel time-to-digital converter (TDC). 

The need for fast feedback has also driven recent hardware development.  A modular FPGA-based system called QubiC is proposed in \cite{xu:2021} for the measurement and control of superconducting qubit systems that support the execution of gate-based quantum algorithms. The prototype system is designed to generate RF pulses to control and measure qubits, and to provide fast feedback control for QEC. It consists of a Xilinx VC707 FPGA and Abaco Systems FMC120 boards with ADC and DAC modules for the generation and detection of intermediate frequency (IF) signals, an RF mixing module for signal conversion and a local oscillator (LO) implemented with a master oscillator driving the inputs of multiple phase-locked loops (PLLs). 

The system described in \cite{gebauer:2020a, gebauer:2021} is also designed to meet the instrumentation requirements of superconducting qubit systems but shares similar platform characteristics to the one that we propose. A modular approach is taken in which the PL side is partitioned into regions called digital unit cells, each of which is responsible for managing one qubit. The APU and RPU components of a ZCU111 are used to implement user interface functions and to provide low-latency run-time configuration and data processing with PL components, respectively. However, the proposed system uses AXI4-Lite memory-mapped register interface to the PL and a 2-to-N Wishbone bus system 
for retrieving data, which can limit its data processing and feedback.

A recently proposed quantum instrumentation control system (QICK) is described in \cite{Stefanazz:2022}. The system utilizes the ZCU111 RFSoC and is capable of controlling multiple qubits with direct synthesis of control pulses with carrier frequencies up to 6 GHz. The programmable logic is configured with a customized module that can synthesize and digitally upconvert arbitrary pulses, measure and downconvert incoming signals, as well as react in real-time to feedback. The system utilizes the APU (which runs python applications under Linux), a timed-processor implemented in the PL, and DMA transfers between the APU and PL, but does not incorporate the RPU. The authors describe the analog performance and digital latency of the system but do not characterize the performance characteristics of the various communication channels within the ZCU111.

Multiple FPGA-based commercial systems have recently become available and include integration with other key hardware.  For instance, Sinara is a hardware control system that uses the open-source ARTIQ software for supporting quantum applications \cite{kasprowicz:2020}.  
The ecosystem offers modules that include AWGs, DDSs, RF generators, and feedback elements such as proportional–integral–derivative (PID) servos, in addition to carrier cards that use FPGAs to coordinate an experiment. Liquid Instruments, Quantum Machines, Zurich Instruments, Keysight, and National Instruments also provide electronic control hardware that is tailored for quantum computing.

Several recent research efforts describe throughput characteristics of MPSoCs outside of the context of quantum computing. An analysis of the throughput and latency of AXI port configurations for data transfers between memory subsystems is presented for the ZCU102 and Ultra96 boards in \cite{vaishnav:2019}. AXI bus widths, burst size, memory chip configuration, access patterns and transaction frequency are taken into consideration in their analysis. The analysis however is presented for DMA transfers using the APU between PS DDR and PL Block RAM (BRAM), and does not address RPU, PL DDR, inter-processor communication and general-purpose input/output (GPIO) performance characteristics. A second investigation of memory performance parameterized by burst and memory stride sizes within the ZCU102 is presented in \cite{argyriou:2021}. The focus of the analysis is again on DMA transfers from PS DDR to PL BRAM.   %Moved content to intro
\section{Architecture}
\label{sec:architecture}

The goal of our work is to define an RFSoC platform configuration that meets the requisite electronic performance (e.g. timing, phase accuracy, amplitude stability) for performing high-fidelity quantum gates while supporting flexible gate sequences and the ability to extend the hardware to control more qubits.  To achieve this we start with these high-level design principles:

\begin{enumerate}
    \item \textbf{RPU vs. APU task division}: assign operations that require strict deadline-based timing to the RPU and PL state machines, whereas overall coordination and tasks with soft real-time-based timing constraints are assigned to the APU.
    \item \textbf{Multiple processors and DMA for increased parallelism}: control and synchronize RPU operations using the APU, and leverage an RPU core for meeting hard real-time deadlines. Maximize hardware parallelism by fully utilizing AXI interconnect between processing cores, and for carrying out high-speed gate-sequence-based data transfers using DMA between the PS/PL DDRs and PL AXI streaming interfaces. 
    \item \textbf{Leverage modern classical computing paradigms}: use optimized commercially-developed hard-wired processing blocks where possible to minimize latency and maximize bandwidth.
\end{enumerate}

Applications running on the APU provide for a high-level language abstraction for carrying out soft real-time-based complex computing tasks, with access to comprehensive library functions, and internet-based access and data transfer mechanisms. Light-weight, real-time bare-metal and FreeRTOS applications running on the RPU connect to both the APU and PL-side components using fast on-chip interconnects for interprocess communication, via Open Asymmetric Multiprocessor (OpenAMP) and RPMsg, GPIO, and block-oriented DMA transfer mechanisms. 
\section{Gate sequence generation and flow}
\label{sec:dds}

The fundamental job of the coherent control system is to compose RF waveforms that implement sequences of high-fidelity quantum gates. In their simplest form, these pulses consist of RF oscillations with a square envelope and defined frequency, phase, and amplitude. Fluctuations in the calibrated values for these control parameters are a common source of gate error.  These fluctuations can be categorized into two general regimes: fast shot-to-shot fluctuations with typically $\ll$1\% relative amplitude, and slow drifts on timescales ranging from seconds to hours that can lead to larger relative errors after long run times. 
%\comment{Dan: what about noise that is faster than the gate time?}  
Shot-to-shot fluctuations can be mitigated using dynamical-decoupling gates \cite{merrill:2014}, such as BB1 or SK1 gates for pulse length errors (PLE), and CORPSE or Q1 gates for off-resonant errors (ORE), or combinations of the two (CCCP or B2CORPSE). These gates often require discrete phase jumps, and in some cases continuous amplitude modulation (AM) (e.g. Q1, Q2, S1, and S2). Dynamical decoupling schemes for two-qubit entangling gates can also require techniques such as continuous frequency modulation (FM) or combinations of FM and AM.
Other state-of-the-art quantum gate designs and pulse engineering techniques, such as GRadient Ascent Pulse Engineering (GRAPE) rely on simultaneous modulation of all parameters.

Supporting continuous modulation across all parameters requires large amounts of data, and such techniques are often implemented using arbitrary waveform generators (AWGs). AWGs suffer from long load times and limited circuit depth, due to the sheer number of points that must be encoded to describe the full waveform. However, even the more advanced dynamical decoupling gates require modulation envelopes with spectral components which have relatively low frequencies in comparison to the baseband RF frequency, at least for trapped ions, and therefore more memory-efficient encodings are possible outside of AWGs. 

\begin{figure*}[htbp]
    \centering
    \includegraphics[width=\linewidth,keepaspectratio=true]{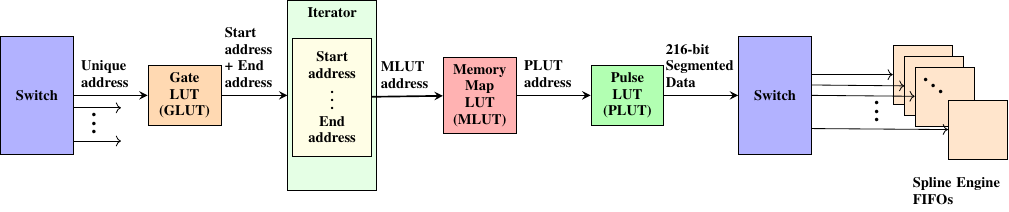}
    \caption{LUT architecture utilized for representing compressed gate sequences. \change{A more detailed description of how gate sequences are generated is provided in the appendix (\ref{sec:gate_sequencers})}.}
    \label{fig:luts}
    \vspace{-2pt}
\end{figure*}

We exploit this disparity by implementing a custom arbitrary waveform modulator (AWM), which supports advanced modulation of waveform control parameters and requires only a fraction of the data required by an AWG. Our AWM consists of two main elements, a custom DDS module and a gate sequencer. 
The DDS module implements global phase synchronization for automatic phase bookkeeping and supports specialized features such as frequency feedforward corrections and dynamic cross-talk cancellation for shimming out external hardware errors. The gate sequencer module is responsible for scheduling waveform parameters that are fed to the DDS.  
These parameters are fed using a hierarchical set of look-up tables (LUTs) shown in Fig. \ref{fig:luts}.  The gate LUTs and memory map LUTs provide memory-efficient reference to specific sequences of pulse information that comprise quantum gates, and are described in the appendix (\ref{sec:appendix_dds}). 
This pulse information is stored at the lowest level in a pulse LUT with 216-bit spline data that parameterizes the frequency, amplitude, phase, and timing of the gate.  These are the values that must be collectively updated and communicated when gates are reconfigured.  Based on the amount of information needed for specifying a gate, we find the communication speeds measured in the next section to be sufficient for supporting reconfiguration times that are less than 2\mics\!, on the order of the fastest gate time.
\section{Communication}
\label{sec:com}

\begin{figure*}
\vspace{-8pt}
   \centering
   % \scriptsize\input{images/ZCU111_interconnect.tikz}
  \includegraphics[width=6.5in,keepaspectratio=true]{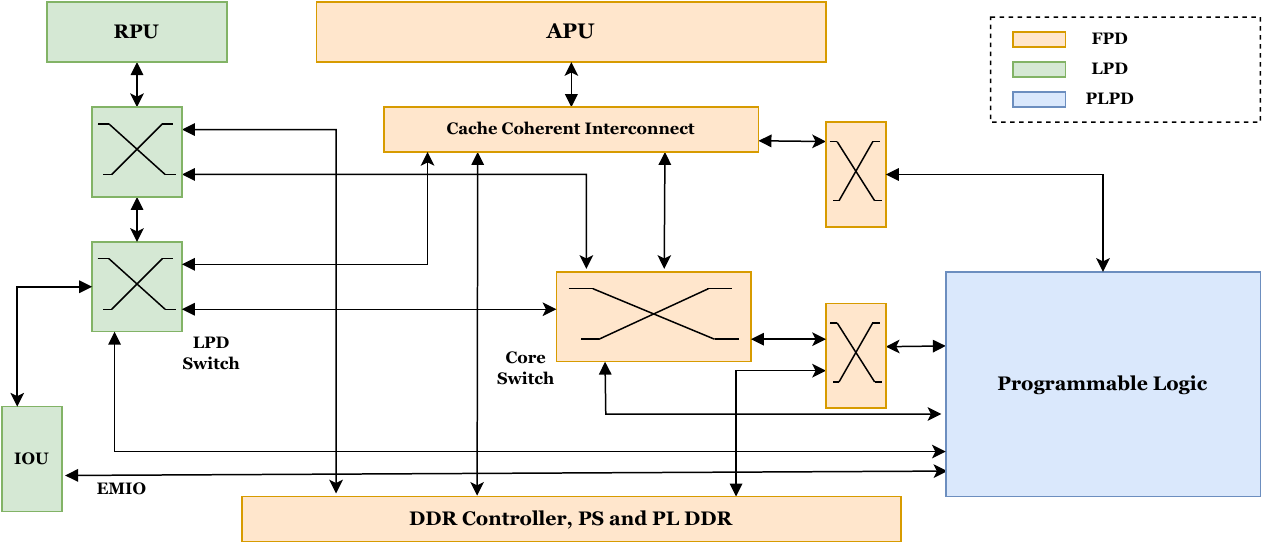}
   \caption{Zynq MPSOC system architecture showing microprocessors, programmable logic, DDR and interconnect.}
   \label{fig:ZCU111_architecture}
   \vspace{-5pt}
\end{figure*}

% \begin{figure*}
% \vspace{-8pt}
%    \centering
%    \includegraphics[width=\linewidth,keepaspectratio=true]{images/irtij3.pdf}
%    \caption{Zynq MPSOC architecture for a TIQC control system.}
%    \label{fig:Zynq_communication_channels}
%    \vspace{-10pt}
% \end{figure*}

\begin{figure*}
\vspace{-8pt}
   \centering
   % \scriptsize\input{images_archive/ArchDiagram.drawio.tikz}
  \includegraphics[width=5.0in,keepaspectratio=true]{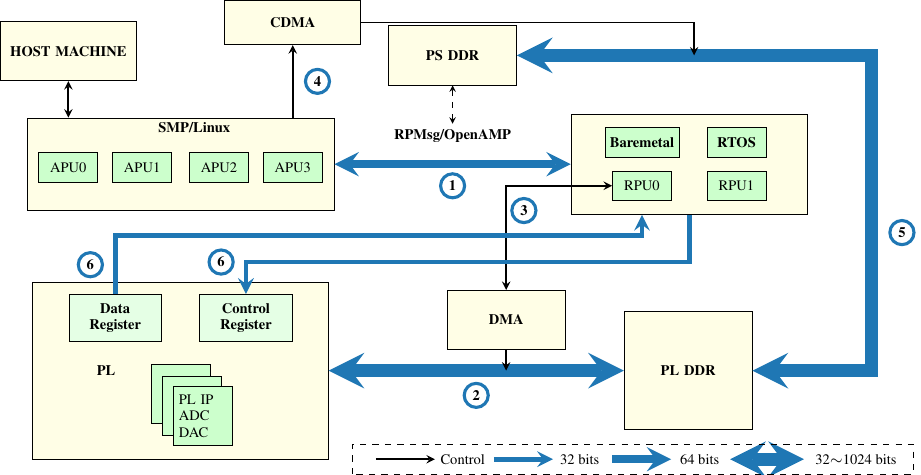}
   \caption{Zynq MPSOC communication channels.}
   \label{fig:Zynq_communication_channels}
   \vspace{-5pt}
\end{figure*}

Having described the gate sequencer LUT architecture and general division of labor between the APU and RPU, in this section we describe the RFSoC (ZCU111) hardware and experimental results related to communication within the system. 
As discussed earlier, the hardware architecture of the ZCU102 and ZCU111 are very similar with respect to the interprocessor communication mechanisms and AXI interconnect architecture, and therefore, only the ZCU111 performance metrics are shown. The latency and throughput measurements for DMA transfers, on the other hand, exhibit significant differences as we show in the following section. 

A block diagram of the processing and interconnect components within the Zynq UltraScale+ SoC on the ZCU111 (and ZCU102) is shown in Fig.~\ref{fig:ZCU111_architecture}. The processing system \change{includes a set of AXI switches that interconnect the five main components of the SoC system architecture, namely, the dual-core Cortex R5 RPU, quad-core Arm Cortex A53 APU, Programmable Logic, DDR and IO unit. Xilinx uses the terms full-power-domain (FPD), low-power-domain (LPD) and programmable-logic-power-domain (PLPD) to refer to regions on the SoC that have separate power control mechanisms, with each referring to the power domains for the APU, RPU and Programmable Logic, respectively. The RPU and APU have} access to a DDR4 4 GB 64-bit PS SDRAM and the 4 GB 64-bit PL DDR4 (the ZCU102 possesses a 512 MB 16-bit PL DDR4 and provides lower performance as we discuss in the following). 

\change{The corresponding communication channels within the system architecture that we evaluate in this paper are shown in Fig. \ref{fig:Zynq_communication_channels}}. The blue arrows illustrate the \change{data transfer paths between the APU, RPU and Programmable Logic, while the black arrows represent the control signals. Each of the communication mechanisms support parallel transfer capability of at least 32 bits. While GPIO and EMIO are limited to 32 bits in our experiments, RPMsg supports 64-bit transfers while DMA and CDMA are variable and can be expanded up to 1024 bits. The thickness of the blue arrows and the legend identify characteristics of the data transfer paths.} 

Communication performance is characterized by two primary parameters, latency and throughput (the term bandwidth is used in reference to the maximum achievable throughput). Both latency and throughput are subject to variation because of interfering events, e.g. interrupt processing by an APU, blocking events within the switches of the interconnect, refresh cycle requirements of the DDRs, and others. It is particularly important to determine both the average value and variability in these parameters, since a quantum computer cannot store quantum states at checkpoints while it pauses for communication, but instead must perform continuous cycles of quantum error correction. 
The standard method of computing average values and variability is to compute the mean and standard deviation, and (1,2,3)$\sigma$ is used to get a sense of confidence intervals given $\sigma$. However, this assumes the variability in the communication mechanisms can be characterized as Gaussian. In our experiments, we rarely found instances of Gaussian behavior\footnote{\change{The most likely reason for the lack of Gaussian behavior is the discrete event-driven characteristic of SoC-based microprocessor systems. Gaussian variations are typically associated with continuous random variables. The time intervals of events such as cache misses, AXI-interconnect blocking events, DDR refresh operations and interrupt service handling are discrete, i.e., are associated with fixed, non-zero time intervals, making the distributions even over large numbers of samples asymmetric and non-uniform. This type of timing behavior justified our use of non-parametric statistics.}} and instead report results using non-parametric statistical metrics, which include the median, minimum, and maximum values. Characterizing the range of variability is especially important for quantum computing where processing of feedback is often time-critical.

The communication mechanisms shown in Fig. \ref{fig:Zynq_communication_channels} are summarized as follows, and described in detail in the following.

%\comment{double check}
\begin{itemize}
    \item RPMsg between the APU and RPU, labeled with \tcirc{1} in the figure.
    \item DMA between a bare metal application running on one of the RPU cores and a streaming AXI interface and state machine in the PL using PL DDR4, labeled \tcirc{2} through \tcirc{3} in the figure.
    \item CDMA between the PS and PL DDR4, controlled by the APU, labeled \tcirc{4} through \tcirc{5} in the figure.
    \item Memory-mapped AXI-Lite GPIO registers and extended multiplexed input-output interface (EMIO) between the RPU and PL, labeled \tcirc{6} in the figure.
\end{itemize}

The communication channels between the APU, RPU, and PL require the configuration and compilation of the custom Linux kernel. Linux is run on top of a symmetric multiprocessing configuration that defines the APU hardware architecture with four cores. The APU and RPU subsystems, on the other hand, define an \textit{asymmetric} multiprocessing (AMP) system, in which the RPU cores operate as independent processor components with respect to the APU. The multi-user, time-sharing system model, which characterizes the Linux OS, is not capable of meeting the deadline-based, real-time system requirements of a quantum computing system. Instead, the RPU is utilized for this purpose, and is configured to run bare metal applications and/or a real-time operating system (RTOS) such as FreeRTOS.

The PL represents a third component that is used for defining highly customized peripherals, that can be utilized for specialized coprocessing tasks. Whereas register-transfer-level or RTL-based design typically involves low-level constructs that lack the expressiveness of high-level software abstractions, it provides an ideal platform for generating custom peripherals where one has absolute control over timing characteristics. Therefore, PL can supplement the PS with features that meet hard, real-time system constraints. 

Each of the illustrated communication channels shown in blue in Fig. \ref{fig:Zynq_communication_channels} utilize a unique interface mechanism. For example, the communication channel between the APU and RPU, labeled as \tcirc{1} in the figure, makes use of an API defined by the libmetal and OpenAMP standards \cite{OpenAMP}, and utilizes on-chip tightly-coupled memory (TCM) for code, stack, heap, etc., and the PS DDR for inter-processor communication (IPC) for data exchange and synchronization. \change{Similarly, the blue arrows labeled \tcirc{2} and \tcirc{5}} capture the sequence of operations that occur during direct memory access (DMA) transfers, whereas the blue arrows labeled \tcirc{6} shows the transfer paths between the RPU and PL when memory-mapped interfaces, including AXI-Lite GPIO registers and EMIO, are used. The architectural details and measurement results for each of these transfer mechanisms are presented in the following subsections.

\subsection{Experimental Setup for GPIO and RPMsg Data Collection and Measurements}

\begin{figure}
    \centering
    \includegraphics[width=\columnwidth,keepaspectratio=true]{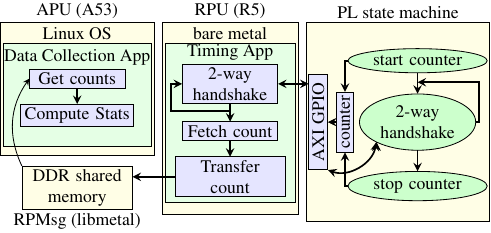}
    \caption{Block diagram showing architecture for timing measurements between the APU, RPU and programmable logic.}
    \label{fig:GPIO_RPU_timing_block_diagram}
    \vspace{-15pt}
\end{figure}

% \begin{figure}
%     \centering
%     \input{images_archive/GPIO_RPU_Timing_block_diagram.tikz}
%     % \includegraphics[width=\columnwidth,keepaspectratio=true]{images/GPIO_RPU_Timing_block_diagram.jpg}
%     \caption{Block diagram showing architecture for timing measurements between the APU, RPU and programmable logic.}
%     \label{fig:GPIO_RPU_timing_block_diagram}
%     \vspace{-15pt}
% \end{figure}

\begin{figure}
    \centering
    \includegraphics[width=\columnwidth,keepaspectratio=true]{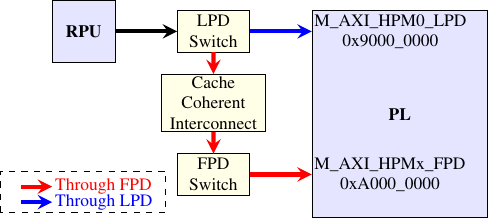}
    \caption{Routing network for two AXI-Lite GPIO configurations.}
    \label{fig:GPIO_Interconnect_block_diagram}
    \vspace{-15pt}
\end{figure}

The Libmetal API provides applications with access to interprocessor interrupts (IPI) and shared memory for instrumenting communications between two or more processing units. The OpenAMP framework builds on top of Libmetal to provide a higher level of abstraction to these communication services, which are referred to as Life Cycle Management (LCM) and Interprocessor Communication (IPC). The Libmetal/OpenAMP API defines a communication mechanism based on remoteproc and RPMsg driver primitives, which are implemented within the Linux kernel and RPU bare-metal application. The remoteproc API allows applications that run on the APU to initialize, start and terminate binary executables on the RPU, whereas the RPMsg API defines a protocol for interprocessor communications. 

Xilinx hardware development tools, including Vivado and Vitis \cite{Vivado}, are used to create the architecture shown in Fig. \ref{fig:GPIO_RPU_timing_block_diagram} for the AXI-Lite GPIO and EMIO latency and throughput measurements. The timing process starts with the APU loading a bare-metal application on the RPU. The APU C program then initializes the RPMsg communication facility between the APU and RPU and transfers run-time parameters to the RPU (not shown). The RPU program receives the configuration parameters and then enters a loop that reads and writes two 32-bit memory-mapped registers as a means of exchanging information with a state machine (SM) running in the PL. The SM instantiates a latency and throughput counter that are used to record the number of PL clock cycles required to execute a handshake communication protocol between the RPU and PL.  

We investigate two AXI-Lite GPIO configurations which map to different physical addresses as shown in Fig.~\ref{fig:GPIO_Interconnect_block_diagram}. Surprisingly, the physical address assigned impacts the performance characteristics. The first configuration, labeled `Through FPD', memory maps the GPIO just above the upper limit of the LPD aperture at 0xA000\_0000 and requires communication traffic to route through both the LPD and FPD switches. The LPD aperture is defined as a 512 MB region between 0x8000\_0000 and 0x9FFF\_FFFF in the RFSoC and MPSoC architectures. The second configuration, labeled 'Through LPD', maps the GPIO into a small region of the LPD physical address aperture at 0x9000\_0000. The RPU communication traffic in this case routes to and from the PL using only the LPD switch. The additional routing in the first configuration increases latency and decreases throughput. 

The primary consideration here for quantum systems is the limited size of the memory region directly accessible by the RPU in the RFSoC and MPSoC system architectures, namely the 512 MB LPD region. The primary data transfer mechanism between the RPU and the PL is DMA, which also needs to utilize this memory region. Moreover, a custom Linux kernel is built to map a portion of the PL DDR address space into this region as a means of fully utilizing the capabilities and capacities of the two DDR memories. Therefore, a tradeoff exists between maximizing DMA transfer buffer size and achieving the best AXI-Lite GPIO performance. EMIO represents a nearly equivalent alternative to AXI-Lite GPIO with regard to performance, but avoids the limited LPD address space problem, as we discuss in the next section.

\subsection{GPIO Experimental Results} \label{sec:GPIOExperimentalResults}

The RPU AXI-Lite GPIO interface is tested at three different PL frequencies, including 100 MHz, 200 MHz and 333 MHz (the maximum allowed). Latency is measured as the average transfer delay of handshake operations between the RPU and PL SM. Handshaking is implemented by toggling two control bits in the RPU-to-PL and PL-to-RPU GPIO registers. The latency counter in the PL measures the round trip delay, and then divides by 2 to obtain the one-way transfer latency. The expression for latency is $N_c/(2 f_{clk})$, with $N_c$ representing the value of the counter in the PL and $f_{clk}$ the PL clock frequency. The SM starts the counter in the same clock as the PL-to-RPU assertion and stops it once it receives the RPU-to-PL acknowledgement. Note that this assumes the latency of the interconnect is symmetrical between the RPU and PL.

The counter value is then transferred to the RPU, which sends the value via RPMsg to the APU. The APU converts the count to nanoseconds and gathers statistics from multiple trials. The throughput measurement begins the same way but continues through multiple handshake exchanges, as specified by the run-time parameters. Moreover, a complete two-way handshake includes two additional busy waits for the GPIO control bits to return to zero. The count values after multiple exchanges are $(N_B N_I N_H f_{clk})/N_c$ B/s. Here, $N_B$ represents the number of bytes per handshake (4 for the 32-bit GPIO registers), $N_I$ is the number of iterations (90 in our experiments) and $N_H$ is the number of handshakes per iteration (4). 

\begin{figure}
    \centering \includegraphics[width=\columnwidth,keepaspectratio=true]{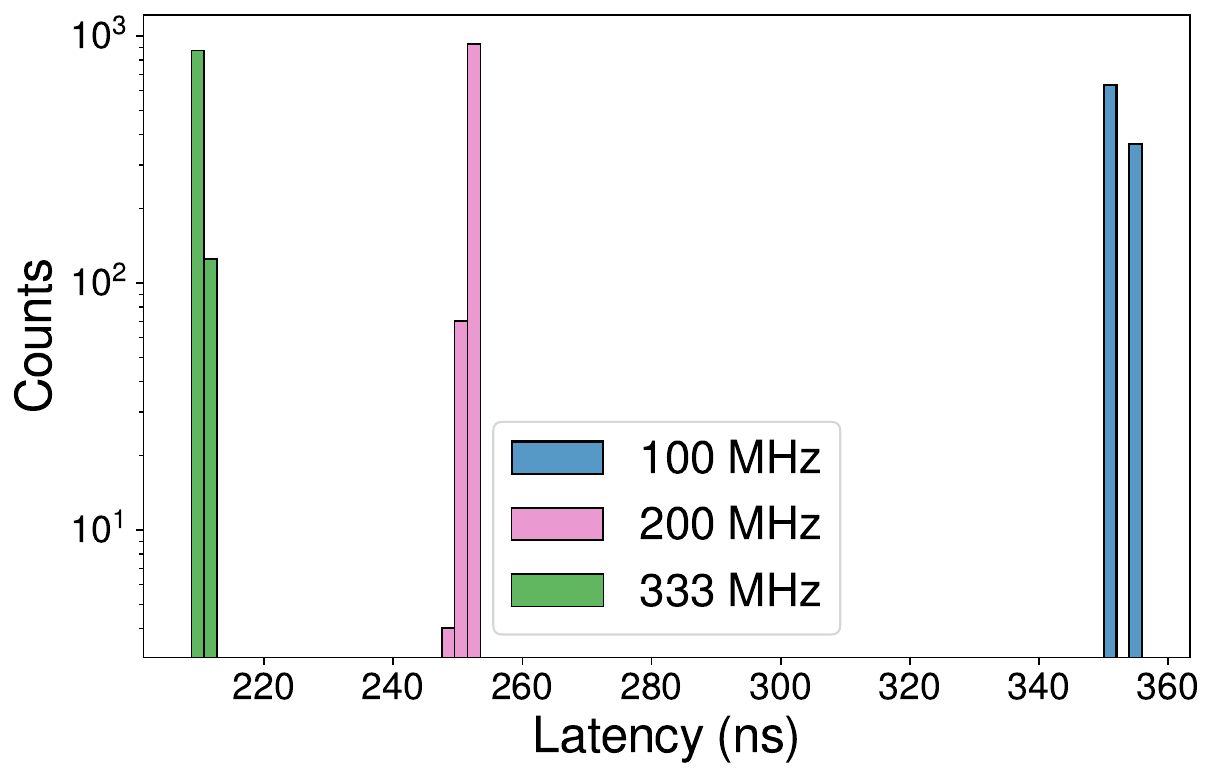}
    \caption{Histogram showing ZCU111 AXI-Lite GPIO latency results for transfers between the RPU and PL using data collected from 1000 individual trials under the `Through FPD' configuration shown in Fig. \ref{fig:GPIO_Interconnect_block_diagram}. Latency results are plotted for test cases with the PL clock frequency set to 100 MHz, 200 MHz and 333 MHz. }
    \label{fig:GPIO_RPU_latency_config1}
    % \vspace{-15pt}
\end{figure}

\begin{figure}
    \centering
    \includegraphics[width=\columnwidth,keepaspectratio=true]{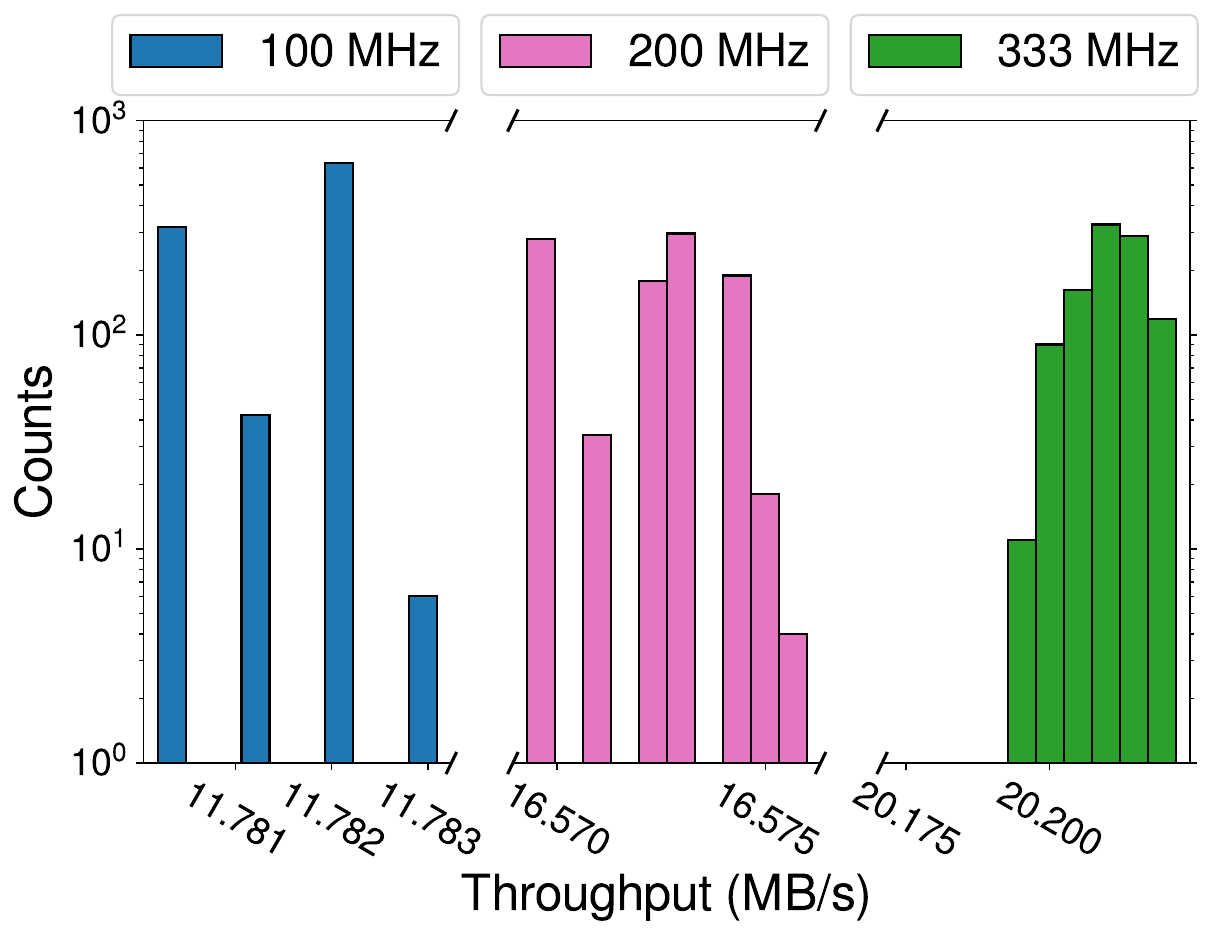}
    \caption{Throughput histogram for AXI-Lite GPIO for the `Through FPD' configuration (companion graph to Fig. \ref{fig:GPIO_RPU_latency_config1}). }
    \label{fig:GPIO_RPU_thrput_config1}
    % \vspace{-15pt}
\end{figure}

\begin{figure}
    \centering   \includegraphics[width=\columnwidth,keepaspectratio=true]{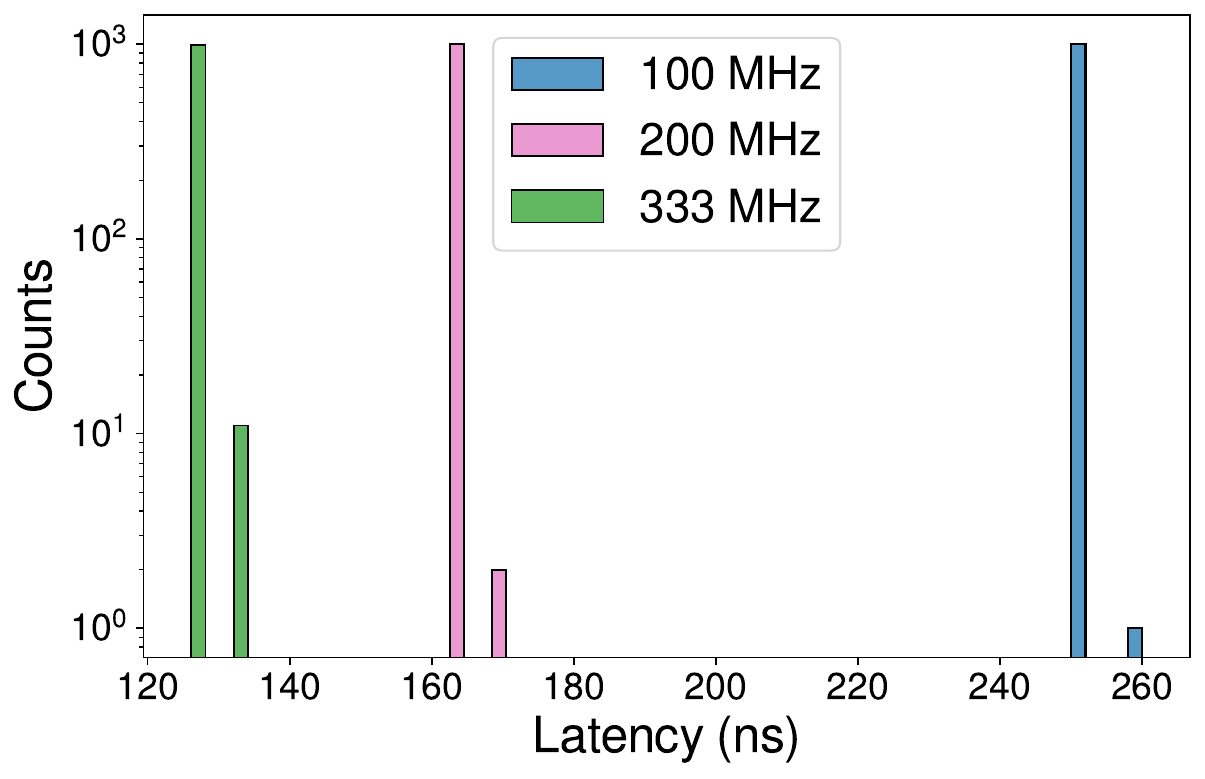}
    \caption{Histogram showing ZCU111 AXI-Lite GPIO latency between the RPU and PL using data collected from 1000 individual trials under the 'Through LPD' configuration shown in Fig.~\ref{fig:GPIO_Interconnect_block_diagram}. }
    \label{fig:GPIO_RPU_latency_config2}
    % \vspace{-15pt}
\end{figure}

\begin{figure}
    \centering
    \includegraphics[width=\columnwidth,keepaspectratio=true]{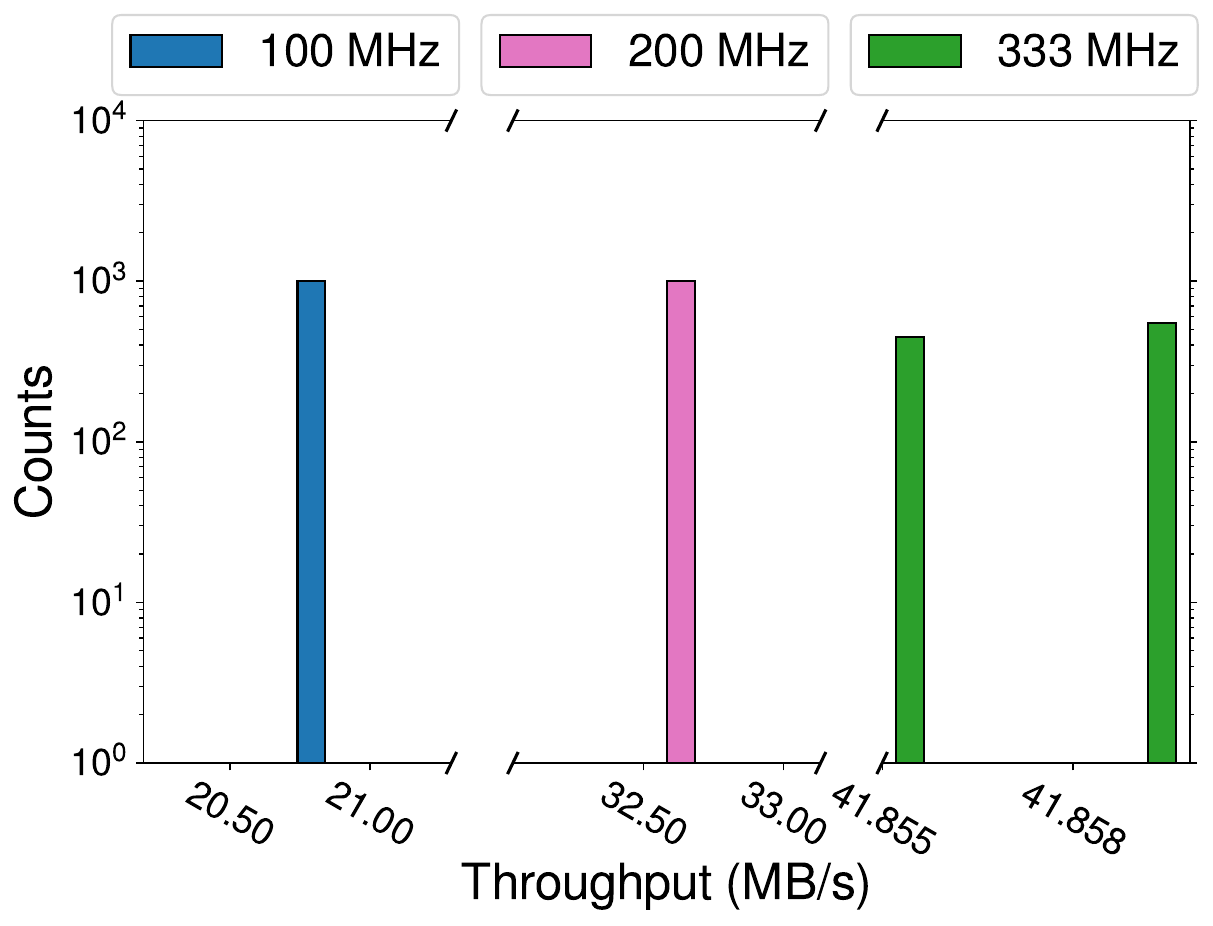}
    \caption{Throughput histogram for AXI-Lite GPIO for the 'Through LPD' configuration (companion graph to Fig. \ref{fig:GPIO_RPU_latency_config2}). }
    \label{fig:GPIO_RPU_thrput_config2}
    % \vspace{-15pt}
\end{figure}

Histograms showing the latency and throughput results from 1000 trials under the `Through FPD' configuration shown in Fig.~\ref{fig:GPIO_Interconnect_block_diagram} are plotted in Figs. \ref{fig:GPIO_RPU_latency_config1} and \ref{fig:GPIO_RPU_thrput_config1}, respectively, for three PL $f_{clk}$ frequencies, 100 MHz, 200 MHz and 333 MHz. The y data is plotted on a $log_{10}$-based scale to enable better visibility of the smaller bin sizes, which portray the variability in the measurements. The median latencies are given as 350, 253 and 210 ns and for throughput as 11.8, 16.6 and 20.2 MB/s, respectively. A second set of histograms are shown for the second AXI-Lite GPIO configuration in Figs. \ref{fig:GPIO_RPU_latency_config2} and \ref{fig:GPIO_RPU_thrput_config2}, with median latencies given by 250, 163 and 126 ns and median throughputs given by 20.7, 32.6 and 41.9 MB/s, respectively. The variation around the median value is less than 15 ns for latency and less than 0.05 MHz for throughput, indicating that GPIO communication is relatively invariant. Note that the GPIO timing measurements are made in a no-load test environment, i.e. the APU is not generating traffic on the AXI interconnect and therefore, these results reflect a best case scenario.

The time spent by the RPU to execute the instructions involving the handshake, and the time spent to transmit data across the AXI-Lite interconnect plus the time required for the PL to consume it, can be calculated from the throughput results. The following simultaneous equations are derived using the median throughput results of the 'Through LPD' experiments. In Eq. \ref{Eq:RPU_AXI_ClkCycles1}, the throughput with the PL configured to run at 333 MHz is converted into the amount of time required to carry out one transfer of 4 bytes.
%/41.9 MB/s $\approx$ 95.6 ns. 
The transaction times in Eqs. \ref{Eq:RPU_AXI_ClkCycles2} and \ref{Eq:RPU_AXI_ClkCycles3} are computed in a similar fashion. Assuming the RPU execution time component $t_{RPU}$ remains constant in all three experiments, and the AXI-Lite interconnect and PL time components scale linearly with frequency, e.g. $t_{PL200} = 1.665 \: t_{PL333}$, two of the following three equations can be solved for the two unknowns and the third equation can be used to validate the results. 

\begin{equation}
    \mathbf{t_{RPU} + t_{PL333} = 95.6 \text{ ns}}
    \label{Eq:RPU_AXI_ClkCycles1}
\end{equation}
\begin{equation}
    \mathbf{t_{RPU} + 1.665 \: t_{PL333} = 122.7 \text{ ns}}
    \label{Eq:RPU_AXI_ClkCycles2}
\end{equation}
\begin{equation}
    \mathbf{t_{RPU} + 3.33 \: t_{PL333} = 193.2 \text{ ns}}
    \label{Eq:RPU_AXI_ClkCycles3}
\end{equation}

The values obtained for $t_{RPU}$ and $t_{PL333}$ are 54.7 ns and 40.9 ns, respectively. An estimate of the error can be computed by subtracting the left-hand-side from the right-hand-side of Eq. \ref{Eq:RPU_AXI_ClkCycles3}, which yields a value of 2.3 ns. Therefore, the RPU runs for 27 clock cycles while the AXI-Lite/PL runs for 14 clock cycles during each transfer operation, with the uncertainty in the estimates equal to only one RPU clock cycle.

EMIO does not utilize the AXI-Lite protocol and is instead structured as a direct multiplexer-based connection network between the RPU and the PL. The RFSoC and MPSoC provide up to 95 configurable single-bit channels that can be used for data transmission in either direction. In our experiments, we configure two 32-bit channels in a fashion identical to the AXI-Lite GPIO, and make latency and throughput measurements using the handshake protocol described earlier. EMIO is limited to a maximum clock frequency of 100 MHz but the PL frequency can be increased to 500 MHz, which represents the configuration used in our experiments.

This particular clock frequency combination also enables latency and throughput to be measured using a one-way transfer mechanism. In the one-way experiments, the C code for the RPU simply toggles a control bit in a tight loop and does not wait for an acknowledgement from the PL. The PL, running at five times the EMIO frequency, over-samples the EMIO control bit to determine the rate at which the bit is toggled by the RPU. Surprisingly, the results from this one-way experiment show that throughput is not symmetric, with RPU-to-PL exhibiting higher throughput than the throughput computed using the two-way handshake configuration where the RPU waits for a PL-to-RPU acknowledgement before executing the next toggle operation. 

\change{Although histograms for the EMIO results are not shown, the median} latency and throughput measured for the one-way experiments are 90 ns and 44.4 MB/s, whereas, \change{for the two-way transfer experiments, the round trip latency degrades} to 370 ns with an average throughput of 20.6 MB/s. \change{This implies that PL-to-RPU latency and throughput are 280 ns and 14.29 MB/s, respectively. The minimum and maximum latencies for RPU-to-PL transfers are 90 and 105 ns, respectively, while the minimum and maximum throughputs are 38.1 and 44.4 MB/s. For PL-to-RPU transfers, the values are 280 and 290 ns, and 13.8 and 14.29 MB/s, respectively.} Given the proprietary nature of the EMIO communication channel, it is difficult to speculate on the reason for the observed asymmetric behavior. 

Overall, the AXI-lite GPIO and EMIO communication mechanisms exhibit low levels of variability, in comparison with several of the other communication mechanisms described in the following sections. This characteristic makes these GPIO-based communication mechanisms attractive for implementing control functions in TIQC systems that have low latency, real-time constraints.

\subsection{RPMsg Experimental Results} \label{sec:RmesgExperimentalResults}

\begin{figure}
    \centering    \includegraphics[width=\columnwidth,keepaspectratio=true]{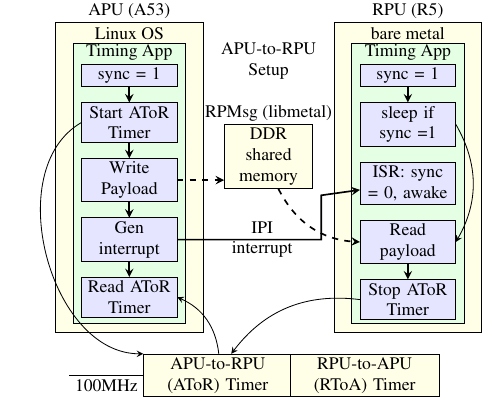}
    \caption{Flow diagram of RPMsg experiment for timing data transfers from APU to RPU. The sequence of operations for measuring throughput from the RPU to APU are identical but reversed.}
    \label{fig:Rmsg_APU_RPU_timing_block_diagram}
    \vspace{-15pt}
\end{figure}

The latency and throughput of messages sent between the APU and the RPU via RPMsg are reported on in this section. A flow diagram that illustrates the process used to make timing measurements is shown in Fig. \ref{fig:Rmsg_APU_RPU_timing_block_diagram}. The APU and RPU both execute a custom timing application under Linux and on bare metal, respectively. Note that the Linux device tree must be configured with elements that support the RPMsg protocol, e.g. shared memory and IPI. Xilinx-provided Vitis software examples are used to create the APU and RPU applications, with a linker script that places the RPU code and data segments into tighly-coupled-memories (TCMs), and shared APU-RPU memory in PS DDR.

% \begin{figure}
%     \centering
%     \input{images_archive/Rmsg_APU_RPU_Timing_block_diagram.tikz}
%     % \includegraphics[width=\columnwidth,keepaspectratio=true]{images/Rmsg_APU_RPU_Timing_block_diagram.jpg}
%     \caption{Flow diagram of Rmsg experiment for timing data transfers from APU to RPU. The sequence of operations for measuring throughput from the RPU to APU are identical but reversed.}
%     \label{fig:Rmsg_APU_RPU_timing_block_diagram}
%     \vspace{-15pt}
% \end{figure}
 
The RPU application, once started by the APU, allocates a block of shared memory and sets up the IPI (not shown). It then loops carrying out the following sequence of operations. A synchronization (semaphore-protected) variable \textit{sync} is initialized to 1 and then the RPU enters a sleep state and waits on an interrupt from the APU. The APU application is then run which also initializes its own \textit{sync} variable and starts the APU-to-RPU (\textit{AToR}) timer.
The APU then writes a payload to the shared memory block in the PS DDR and sends an IPI to the RPU. The RPU's interrupt service routine awakens the RPU application, sets the \textit{sync} variable to 0 
and then reads the payload from shared memory. Once the read operation completes, it stops the \textit{AToR} timer. Although not shown in Fig. \ref{fig:Rmsg_APU_RPU_timing_block_diagram}, the exact same sequence occurs in reverse using a second \textit{RPU-to-APU} timer with the RPU writing a payload to shared memory and the APU performing a read-out. The APU's timing application reads the values of the two timers and stores the values in an array and later to a file for post-processing.

\begin{figure}
    \centering   \includegraphics[width=\columnwidth,keepaspectratio=true]{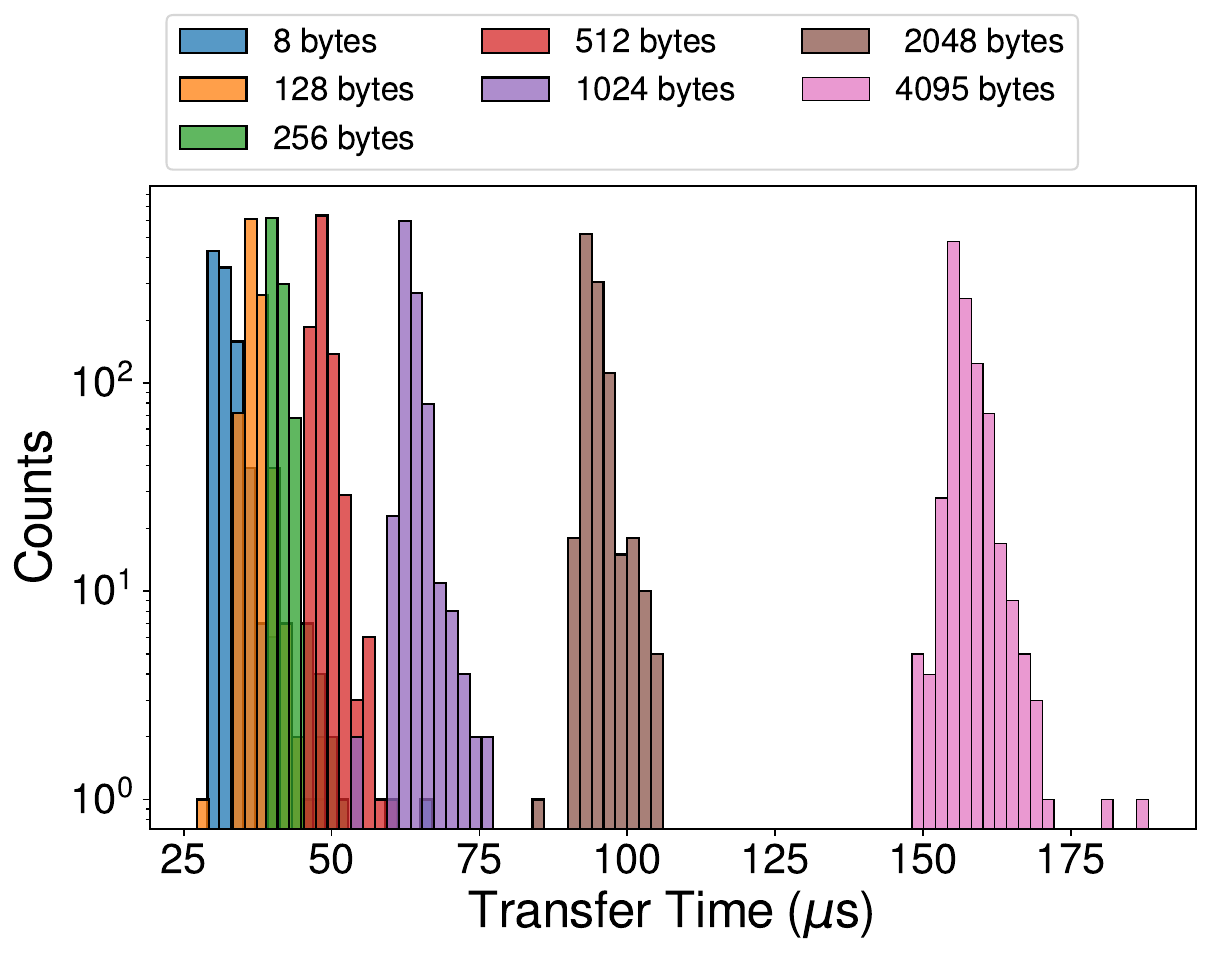}
    \caption{Histogram depicting RPMsg transfer times in microseconds for transfers from the APU to the RPU on the ZCU111. The y data is plotted on a $log_{10}$-based scale to better emphasize variability in the measurements. }
    \label{fig:Rmsg_AToR_TT_results}
    \vspace{-5pt}
\end{figure}

\begin{figure}
    \centering  \includegraphics[width=\columnwidth,keepaspectratio=true]{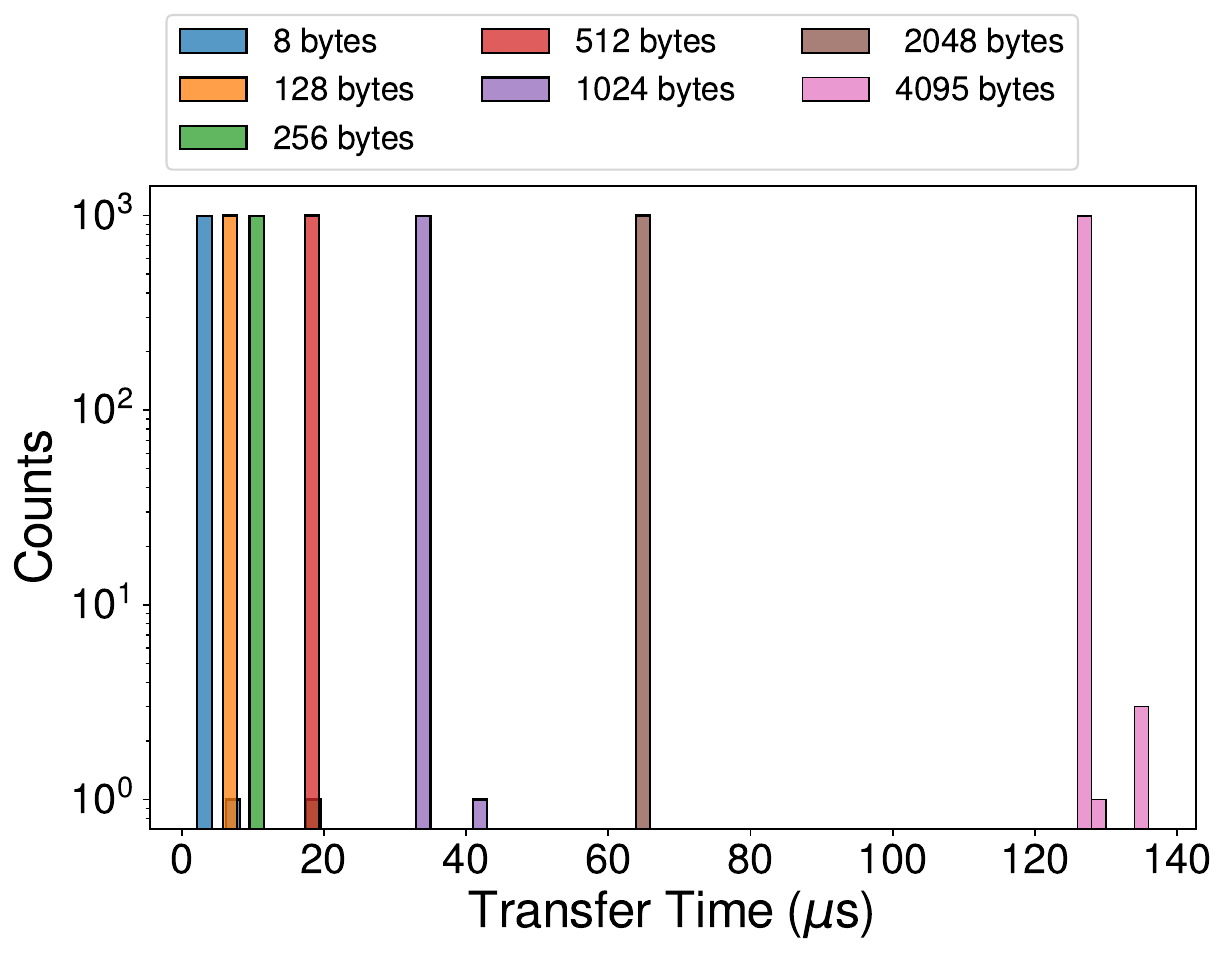}
    \caption{Histogram depicting RPMsg transfer times in microseconds for transfers from the RPU to the APU on the ZCU111. }
    \label{fig:Rmsg_RToA_TT_results}
    \vspace{-5pt}
\end{figure}

% \begin{figure}
%     \centering
%     \includegraphics[width=\columnwidth,keepaspectratio=true]{images/irtij13.pdf}
%     \caption{RPMsg throughput on the ZCU111 showing the median, min, and max throughput characteristics using measurements from 100,000 trials for transfers in both directions, i.e. from APU to RPU (blue) and from RPU to APU (brown). \change{Nafis: Check data for min for APU-to-RPU.}}
%     \label{fig:Rmsg_APU_RPU_throughput_results}
%     \vspace{-5pt}
% \end{figure}

\begin{figure}
    \centering
    \includegraphics[width=\columnwidth,keepaspectratio=true]{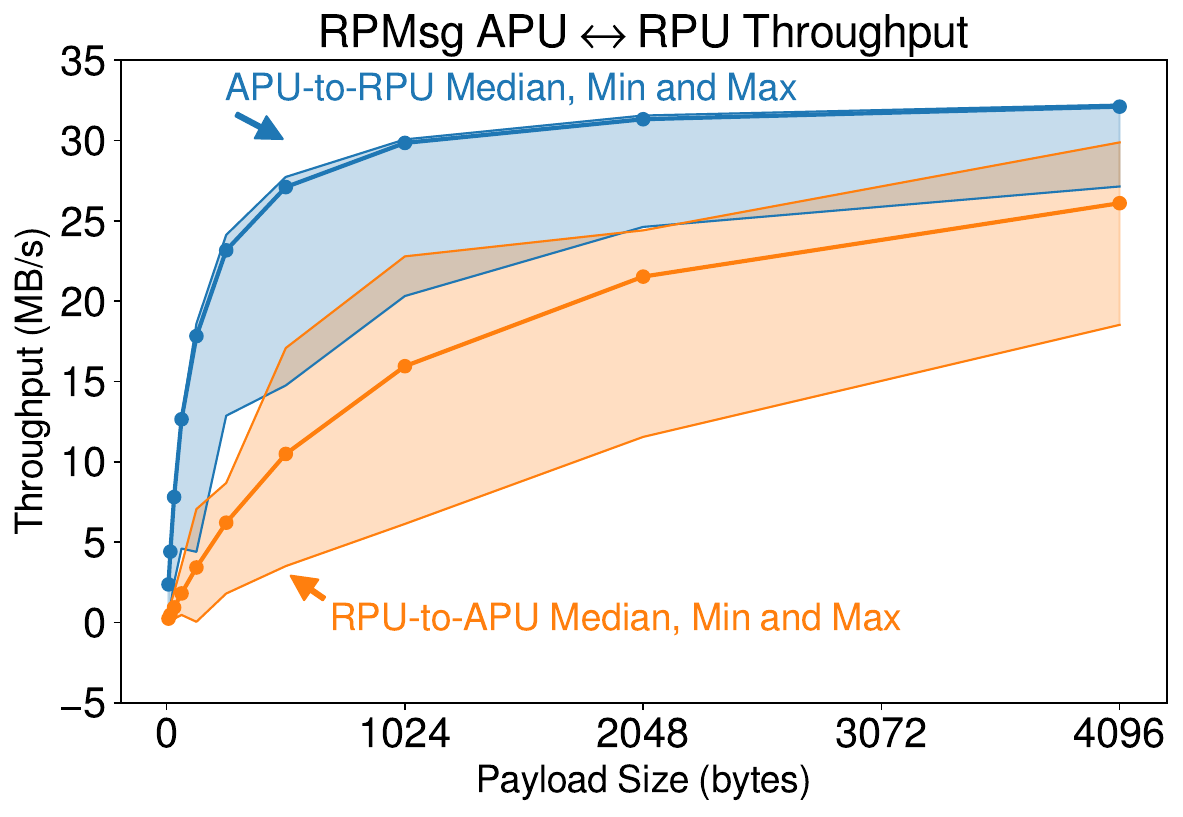}
    \caption{RPMsg throughput on the ZCU111 showing the median, min, and max throughput characteristics using measurements from 100,000 trials for transfers in both directions, i.e. from APU to RPU (blue) and from RPU to APU (brown).% \change{Nafis: Check data for min for APU-to-RPU.}
    }
    \label{fig:Rmsg_APU_RPU_throughput_results}
    \vspace{-5pt}
\end{figure}
The timing application algorithm is repeated using 100,000 iterations for each payload size from 8 bytes to 4096 bytes, with each payload size in the sequence larger than the previous payload by a power of two. The values stored in the arrays are the count values read from the two TTCs, which run at 100 MHz in the PL. Therefore, each count increment represents a time interval of 10 ns. Note that unlike the GPIO measurement scheme, where it was possible to measure latency independent of throughput, it is not possible to determine when the first word of the payload is written to DDR for RPMsg. Given the data bus width of the PS DDR is 8 bytes, we use the 8-byte payload for the latency measurement.

Histograms showing the transfer times associated with the first 1000 trials for a subset of the payload sizes are plotted in Figs. \ref{fig:Rmsg_AToR_TT_results} and \ref{fig:Rmsg_RToA_TT_results} for APU-to-RPU and RPU-to-APU transfer operations, respectively. A key objective here is to portray the asymmetry that exists in the variability of the throughputs for equal-sized payloads in both directions, which is captured well using only subsets of the data. The conversion from TTC counts to transfer time (TT) is given by $TT_{RPMsg}=TTC_{cnt}/f_{clk}$.
Median transfer times vary between 2.2\mics and 126\mics for APU-to-RPU transfers, and 29\mics to 160\mics for RPU-to-APU transfers. 

RPMsg throughput is plotted in Fig. \ref{fig:Rmsg_APU_RPU_throughput_results}, with a maximum rate of 32 MB/s. The RPU-to-APU requires larger payload sizes (not shown) to achieve this throughput rate. \change{Given the high levels of variability in latency and throughput of the RPMsg data transfer mechanism, the qubit control system will utilize RPMsg only for non-real-time operations, e.g. periodic status messages reporting data transfer statistics, debug and error information. In contrast, the much smaller levels of variability associated with the GPIO, EMIO and DMA transfer mechanisms are better suited for qubit data transfer operations that have hard, real-time constraints.}

\subsection{DMA: PL DDR to PL Streaming}
The ZCU102 and ZCU111 support several types of DMA transfer mechanisms. The primary datapath within the qubit system that requires high-speed, block-level transfers is between the PL-side DDR and a streaming interface in the PL (labeled \tcirc{2} through \tcirc{3} in Fig. \ref{fig:Zynq_communication_channels}). A high-bandwidth mechanism to provide updates to gate sequences is critical to tuning and optimizing gate execution as discussed in previous sections. The hard real-time capabilities of the RPU are needed for meeting gate-sequence data transfer requirements and for providing low variability in the response times across multiple, sequential DMA transfers.

\begin{figure}
    \centering
    \includegraphics[width=\columnwidth,keepaspectratio=true]{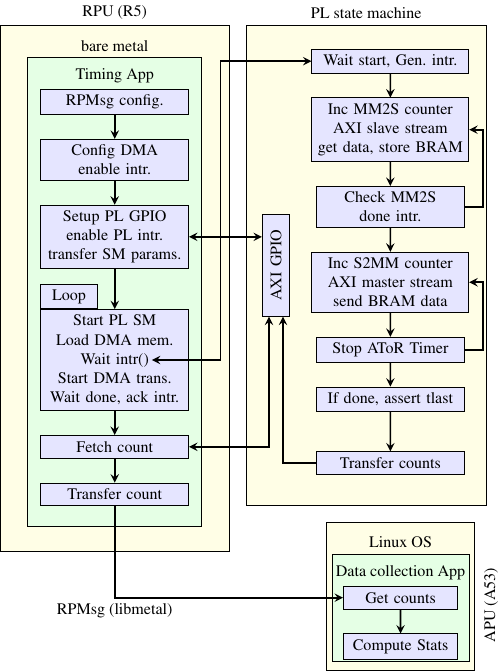}
    \caption{Flow diagram for DMA transfers between the RPU and an AXIS interface implemented in a PL state machine. The RPU is configured to use PL DDR for the data transfers.}
    \label{fig:DMA_RPU_To_PL_Stream_block_diagram}
    \vspace{-5pt}
\end{figure}

A flow diagram of the test procedure is shown in Fig. \ref{fig:DMA_RPU_To_PL_Stream_block_diagram}. Similar to the GPIO experimental setup, the architecture includes an APU, RPU, and a PL SM component, with the APU providing data fetching, analysis, and storage functions only. The RPU performs a sequence of initialization operations related to RPMsg, DMA, and GPIO, including enabling interrupts for the memory-map-to-stream (MM2S) and stream-to-memory-map (S2MM) DMA engine components, and PL interrupts. The RPU transfers parameters to the PL SM, including payload size and parameters for controlling auto-generated interrupts. The auto-generated interrupts are utilized by the measurement system to enable multiple trials to be run back-to-back. 

% \begin{figure}
%     \centering
%     \includegraphics[width=\columnwidth]{images_archive/DMA_RPU_To_PL_Stream_block_diagram.tikz}
%     % \includegraphics[width=\columnwidth,keepaspectratio=true]{images/DMA_RPU_To_PL_Stream_block_diagram.jpg}
%     \caption{Flow diagram for DMA transfers between the RPU and an AXIS interface implemented in a PL state machine. The RPU is configured to use PL DDR for the data transfers.}
%     \label{fig:DMA_RPU_To_PL_Stream_block_diagram}
%     \vspace{-5pt}
% \end{figure}

The RPU carries out multiple, repeated trials for the APU-specified payload size, annotated as Loop in Fig. \ref{fig:DMA_RPU_To_PL_Stream_block_diagram}. In each iteration, the RPU starts the PL SM and blocks waiting for a PL interrupt. The PL SM sends an interrupt to the RPU to start the DMA MM2S transfer operation and simultaneously starts incrementing counters used to determine latency and throughput. The PL SM implements the AXI slave streaming protocol (AXIS) to receive the DMA burst transfers from the RPU. The latency counter runs until the first AXI tvalid assertion occurs while the throughput counter runs until the DMA engine generates a MM2S interrupt done signal. The DMA data block received by the PL SM through the AXIS interface is stored in PL BRAM. The data block is transferred back to the RPU and validated against the original data during the reverse S2MM DMA operation.

The S2MM data transfer operation commences immediately following the MM2S. A third counter is used to determine the S2MM throughput, which measures the time interval between the completion of the MM2S transfer operation and the assertion of the S2MM interrupt done signal. The S2MM DMA channel is notified that the data transfer has completed with the PL SM asserting the tlast signal concurrent with the transfer of the last data word in the payload. The RPU busy waits for the DMA MM2S and S2MM channels to return to an idle state before acknowledging the DMA interrupts. The PL SM transfers the three counter values to the RPU, which forwards them to the APU via RPMsg. This sequence of operations is repeated for each of the trials as specified by the APU.

The APU collects the MM2S latency and throughput and S2MM throughput counter values, and computes the median-min-max statistics using the counts from 10,000 separate trials. The results are stored to a file and later transferred to a host computer. The entire experiment is repeated using PL bitstreams configured with a clock frequency of 333 MHz for the ZCU111 and 300 MHz for the ZCU102, and for payload sizes between 4 B and 1 MB, increasing by powers of two. 

The ZCU102 and ZCU111 integrate different PL DDR memories and therefore, we repeat the experiment above for both and compare the results. The ZCU102 utilizes a 512 MB DDR with a data width of 16-bits whereas the ZCU111 utilizes a 4 GB, 64-bit wide DDR. Therefore, the performance is expected to be higher for the ZCU111, as we show in the following.

% \begin{figure}
%     \centering
%     \includegraphics[width=\columnwidth,keepaspectratio=true]{images/irtij15.pdf}
%     \caption{DMA MM2S throughput for the ZCU111 PL running at 333 MHz and the ZCU102 PL running at 300 MHz for payload sizes of 32, 64 and 128 B, and with the DMA data-bus bit-width set to 256. }
%     \label{fig:DMA_MM2S_thrput_raw}
%     \vspace{-0pt}
% \end{figure}

\begin{figure}[!t]
\centering
\begin{subfigure}[t]{\columnwidth}
\centering
\includegraphics[width=\columnwidth,keepaspectratio=true]{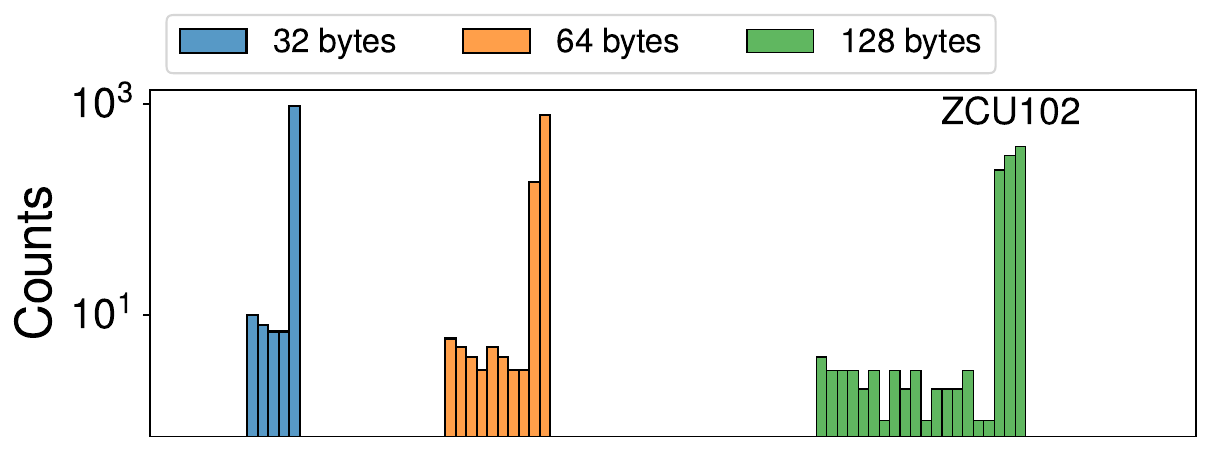}
\phantomsubcaption
\end{subfigure}
\begin{subfigure}[t]{\columnwidth}
\centering
\includegraphics[width=\columnwidth,keepaspectratio=true]{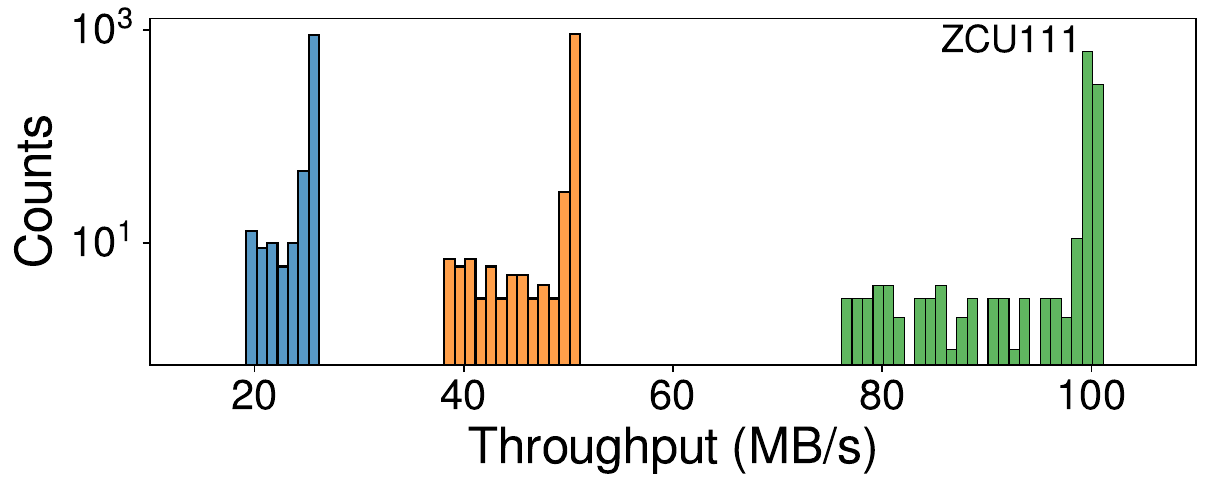}
\phantomsubcaption
\end{subfigure}
\caption{DMA MM2S throughput for the ZCU111 PL running at 333 MHz and the ZCU102 PL running at 300 MHz for payload sizes of 32, 64 and 128 B, and with the DMA data-bus bit-width set to 256. }
\label{fig:DMA_MM2S_thrput_raw}
\vspace{-0pt}
\end{figure}

Latency is explicitly measured for only the MM2S transfer operation because the PL SM is able to measure the time interval between the start of the transfer and the occurrence of the first assertion of the tvalid signal from the MM2S AXI interface. On the other hand, only the start event is known during the S2MM transfer operation, i.e. the PL is not able to determine when the DMA engine successfully transfers the first word into the PL DDR. The PL AXIS interface continuously streams data into the S2MM channel of the DMA engine, which buffers the data internally. For large transfers, it eventually introduces pauses in the PL S2MM AXIS interface because the internal buffer fills up, but the transfer duration for the first word is always measured by our SM as just one clock cycle.

Instead, latency for the S2MM is measured using the throughput counter value with a payload size set to one word, e.g. a 32-byte payload with the DMA configured with a width of 256 bits. The latency measurement for MM2S includes some additional RPU overhead whereas the S2MM does not. This occurs because the PL SM starts the latency counter one cycle after the interrupt is generated, whereas the RPU is blocked waiting for this interrupt. The overhead for the MM2S includes the additional time taken to process the interrupt and to write the length register of the MM2S DMA engine, which effectively starts the MM2S DMA engine. On the other hand, the S2MM is started in advance of the interrupt but blocks until the PL SM reaches the S2MM state. As a consequence, the S2MM latency is much smaller because the RPU overhead does not exist. We report both the MM2S and S2MM latencies recognizing that the true latency is better estimated using the MM2S measurement because the actions required to start the DMA engine are needed in any realistic application scenario.

% \begin{figure}
%     \centering
%     \includegraphics[width=\columnwidth,keepaspectratio=true]{images/irtij16.pdf}
%     \caption{DMA S2MM throughput for the ZCU111 and ZCU102 for payload sizes of 32, 64 and 128 B. }
%     \label{fig:DMA_S2MM_thrput_raw}
%     \vspace{-0pt}
% \end{figure}

\begin{figure}[!t]
\centering
\begin{subfigure}[t]{\columnwidth}
\centering
\includegraphics[width=\columnwidth,keepaspectratio=true]{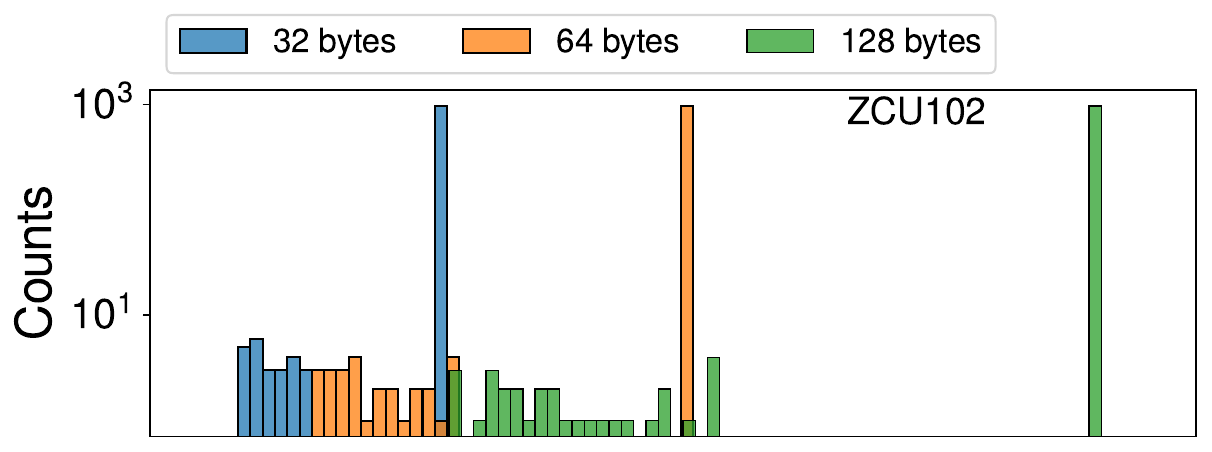}
\phantomsubcaption
\end{subfigure}
\begin{subfigure}[t]{\columnwidth}
\centering
\includegraphics[width=\columnwidth,keepaspectratio=true]{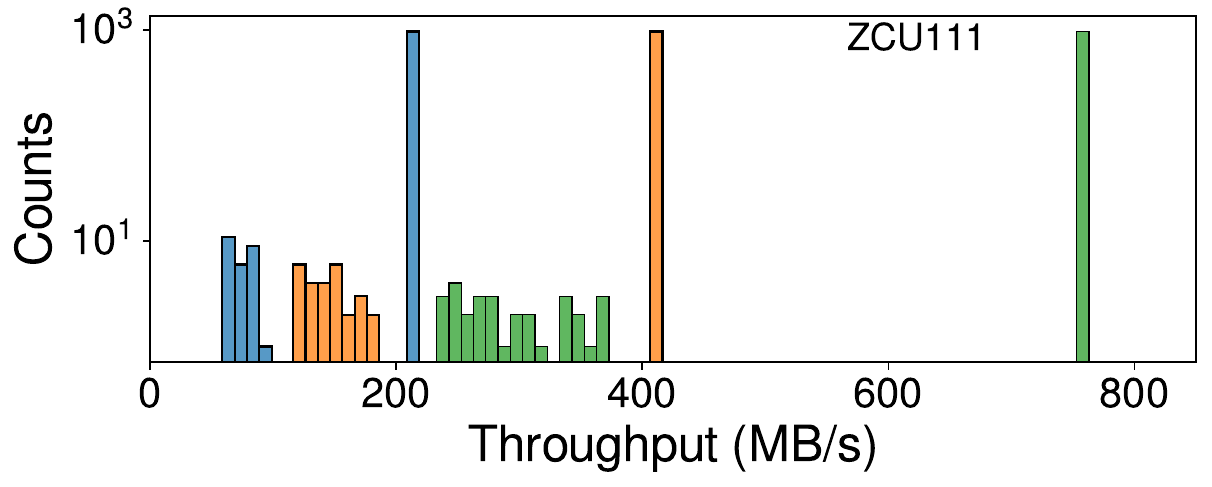}
\phantomsubcaption
\end{subfigure}
\caption{DMA S2MM throughput for the ZCU111 and ZCU102 for payload sizes of 32, 64 and 128 B. }
\label{fig:DMA_S2MM_thrput_raw}
\vspace{-0pt}
\end{figure}

The median-min-max statistics for latency and throughput are computed using equations similar to those given for GPIO. As an illustration, histograms portraying the MM2S and S2MM throughput behavior using data from the first 1000 trials are shown in Figs. \ref{fig:DMA_MM2S_thrput_raw} and \ref{fig:DMA_S2MM_thrput_raw}. The PL logic is configured to run at 300 MHz on the ZCU102 and 333 MHz on the ZCU111, which matches the frequency of the PL-side DDR memories for these devices. The DMA engine is configured with a data bus width of 256 bits and is tasked with transferring payloads of size 32, 64 and 128 B. The individual trials are run back-to-back with approximately 0.5 seconds between trials.

Although the RPU provides un-interrupted execution of the binary program stored in the tightly-coupled memory (TCM), the throughput rates are not constant as one might expect. A periodic decrease occurs in both the MM2S and S2MM throughputs, that is likely due to stalls within the memory interface generator (MIG) to carry out periodic refresh operations. 

\begin{figure}
    \centering
    \includegraphics[width=\columnwidth,keepaspectratio=true]{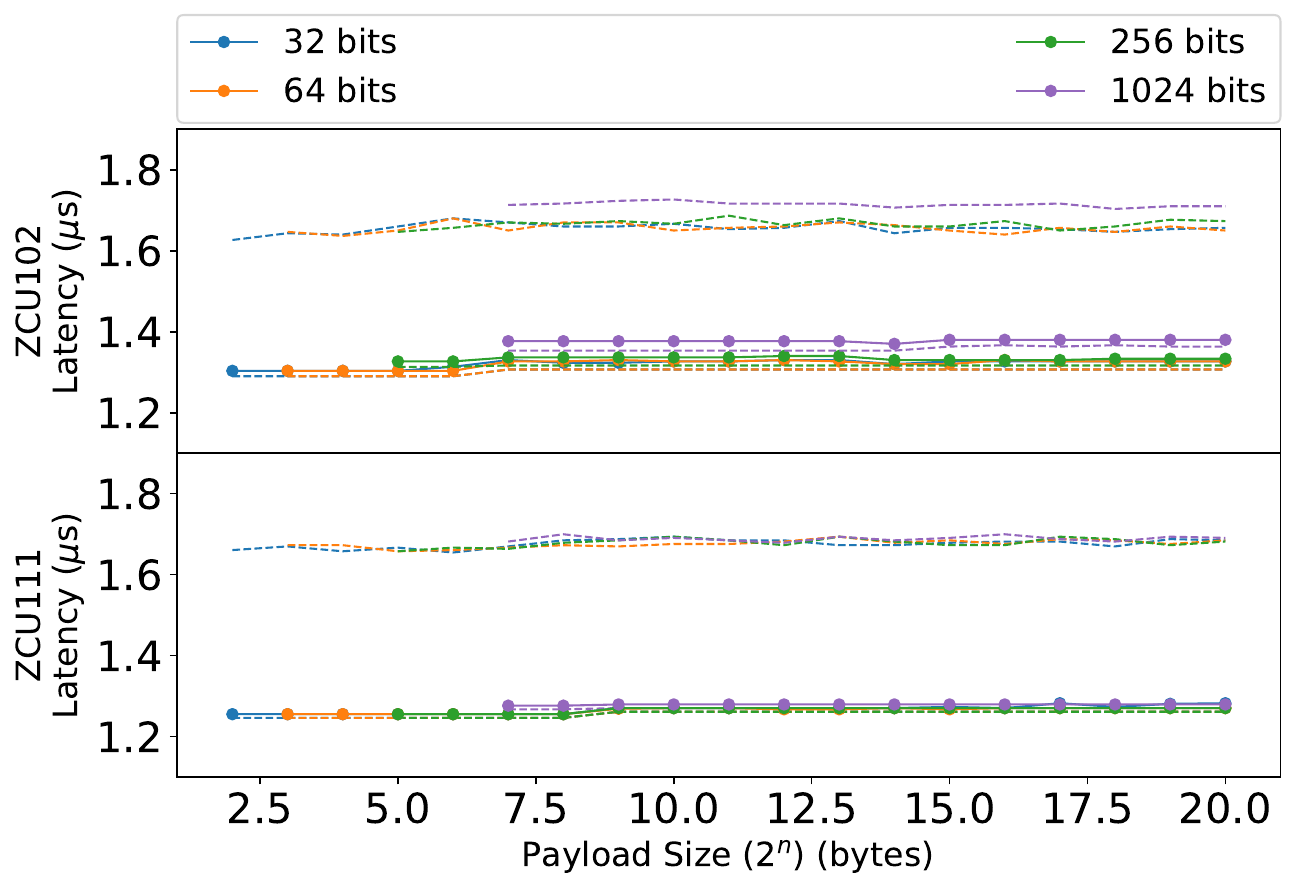}
    \caption{DMA MM2S latency results for the ZCU111 PL running at 333 MHz and ZCU102 PL running at 300 MHz for payload sizes from 4 B to 1 MB, and for DMA data-bus bit-widths of 32, 64, 256 and 1024. The curves defining the medians are color-coded and delineated by the minimum and maximum latency curves shown in red.}
    \label{fig:DMA_MM2S_latency_med_min_max}
    \vspace{0pt}
\end{figure}

% \begin{figure}[!t]
% \centering
% \begin{subfigure}[t]{\columnwidth}
% \centering
% \includegraphics[width=\columnwidth,keepaspectratio=true]{images/18a.pdf}
% \phantomsubcaption
% \end{subfigure}
% \begin{subfigure}[t]{\columnwidth}
% \centering
% \includegraphics[width=\columnwidth,keepaspectratio=true]{images/18b.pdf}
% \phantomsubcaption
% \end{subfigure}
% \caption{DMA MM2S latency results for the ZCU111 PL running at 333 MHz and ZCU102 PL running at 300 MHz for payload sizes from 4 B to 1 MB, and for DMA data-bus bit-widths of 32, 64, 256 and 1024. The curves defining the medians are color-coded and delineated by the minimum and maximum latency curves shown in red.}
% \label{fig:DMA_MM2S_latency_med_min_max}
% \vspace{-0pt}
% \end{figure}

\begin{figure}
    \centering
    \includegraphics[width=\columnwidth,keepaspectratio=true]{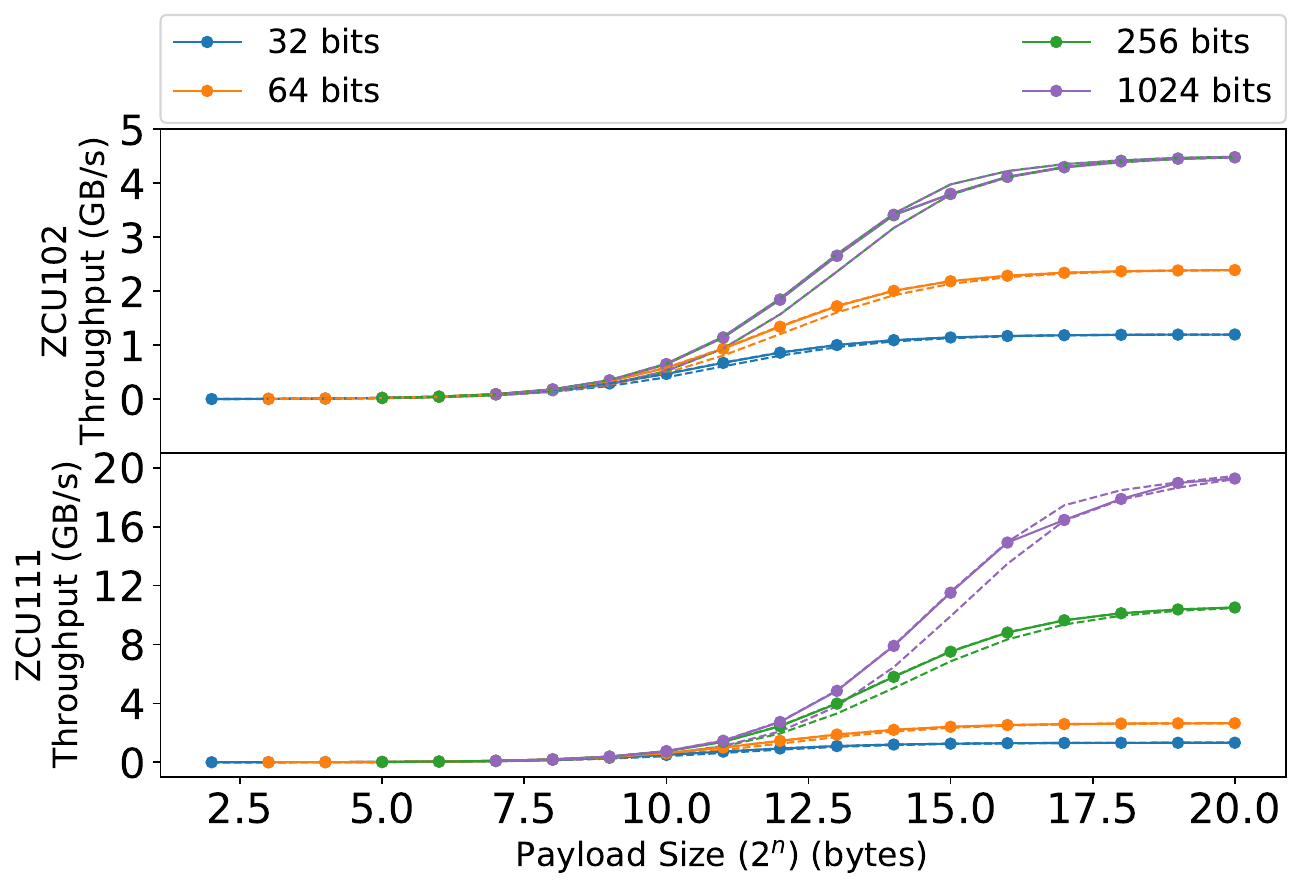}
    \caption{DMA MM2S throughput results for the ZCU111 PL running at 333 MHz and ZCU102 PL running at 300 MHz for payload sizes of 4 B to 1 MB, and for DMA data-bus bit-widths of 32, 64, 256 and 1024. The curves defining the medians pass through shaded regions delineated by the minimum and maximum latency measurements.}
    \label{fig:DMA_MM2S_thrput_med_min_max}
    \vspace{-5pt}
\end{figure}

% \begin{figure}[!t]
% \centering
% \begin{subfigure}[t]{\columnwidth}
% \centering
% \includegraphics[width=\columnwidth,keepaspectratio=true]{images/19a.pdf}
% \phantomsubcaption
% \end{subfigure}
% \begin{subfigure}[t]{\columnwidth}
% \centering
% \includegraphics[width=\columnwidth,keepaspectratio=true]{images/19b.pdf}
% \phantomsubcaption
% \end{subfigure}
% \caption{DMA MM2S throughput results for the ZCU111 PL running at 333 MHz and ZCU102 PL running at 300 MHz for payload sizes of 4 B to 1 MB, and for DMA data-bus bit-widths of 32, 64, 256 and 1024. The curves defining the medians pass through shaded regions delineated by the minimum and maximum latency measurements.}
% \label{fig:DMA_MM2S_thrput_med_min_max}
% \vspace{-0pt}
% \end{figure}

\begin{figure}
    \centering
    \includegraphics[width=\columnwidth,keepaspectratio=true]{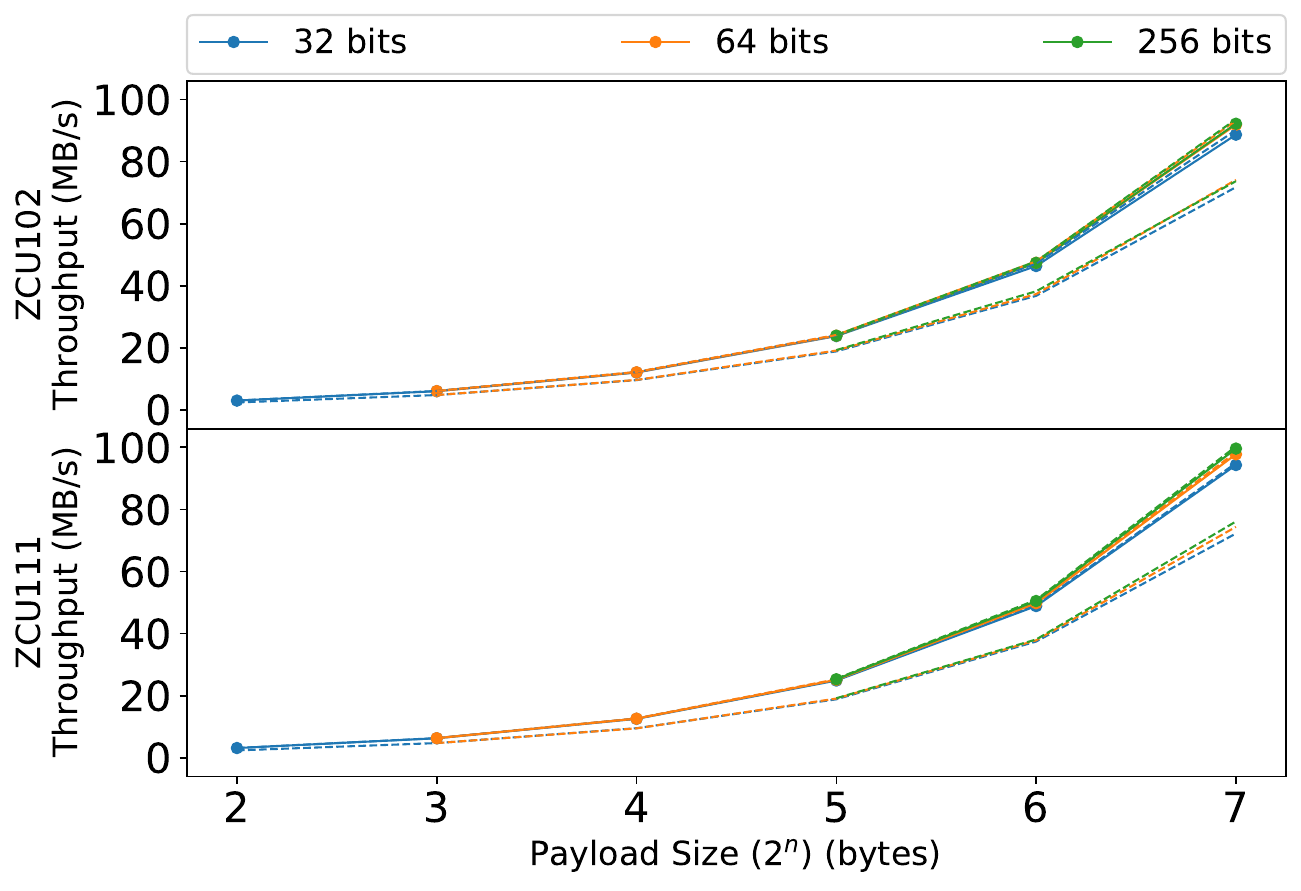}
    \caption{Blow-up of the DMA MM2S throughput results from the left-hand side of Fig. \ref{fig:DMA_MM2S_thrput_med_min_max} emphasizing behavior for smaller payload sizes.}
    \label{fig:DMA_MM2S_thrput_med_min_max_zoom}
    \vspace{-5pt}
\end{figure}

% \begin{figure}[!t]
% \centering
% \begin{subfigure}[t]{\columnwidth}
% \centering
% \includegraphics[width=\columnwidth,keepaspectratio=true]{images/20a.pdf}
% \phantomsubcaption
% \end{subfigure}
% \begin{subfigure}[t]{\columnwidth}
% \centering
% \includegraphics[width=\columnwidth,keepaspectratio=true]{images/20b.pdf}
% \phantomsubcaption
% \end{subfigure}
% \caption{Blow-up of the DMA MM2S throughput results from the left-hand side of Fig. \ref{fig:DMA_MM2S_thrput_med_min_max} emphasizing behavior for smaller payload sizes.}
% \label{fig:DMA_MM2S_thrput_med_min_max_zoom}
% \vspace{-0pt}
% \end{figure}

\begin{figure}
    \centering
    \includegraphics[width=\columnwidth,keepaspectratio=true]{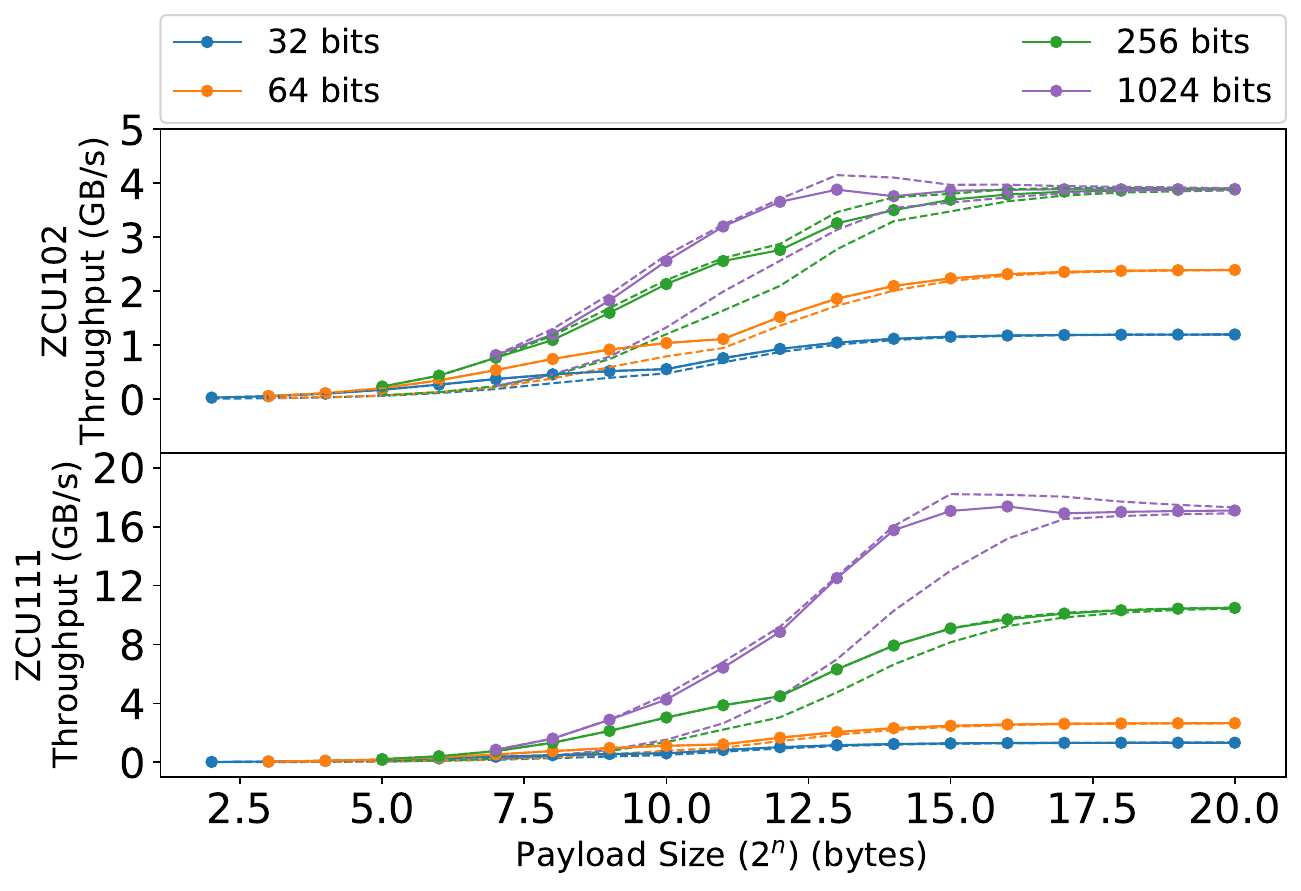}
    \caption{DMA S2MM throughput results for ZCU102 and ZCU111, in the same format as Fig. \ref{fig:DMA_MM2S_latency_med_min_max}.}
    \label{fig:DMA_S2MM_thrput_med_min_max}
    \vspace{-5pt}
\end{figure}

% \begin{figure}[!t]
% \centering
% \begin{subfigure}[t]{\columnwidth}
% \centering
% \includegraphics[width=\columnwidth,keepaspectratio=true]{images/21a.pdf}
% \phantomsubcaption
% \end{subfigure}
% \begin{subfigure}[t]{\columnwidth}
% \centering
% \includegraphics[width=\columnwidth,keepaspectratio=true]{images/21b.pdf}
% \phantomsubcaption
% \end{subfigure}
% \caption{DMA S2MM throughput results for ZCU102 and ZCU111, in the same format as Fig. \ref{fig:DMA_MM2S_latency_med_min_max}.}
% \label{fig:DMA_S2MM_thrput_med_min_max}
% \vspace{-0pt}
% \end{figure}

\begin{figure}
    \centering
    \includegraphics[width=\columnwidth,keepaspectratio=true]{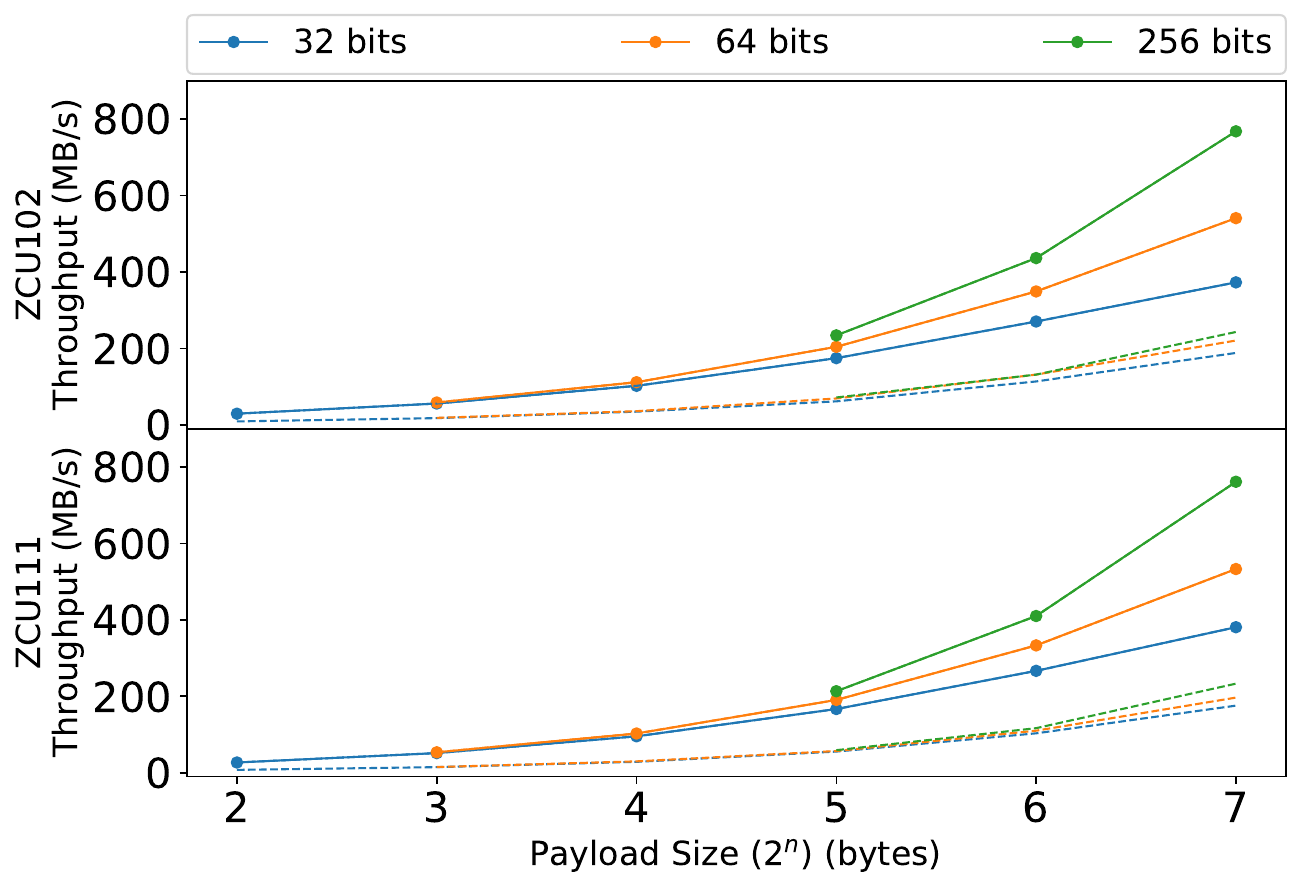}
    \caption{Blow-up of the DMA MM2S throughput results from the left-hand side of Fig. \ref{fig:DMA_S2MM_thrput_med_min_max} emphasizing behavior for smaller payload sizes.}
    \label{fig:DMA_S2MM_thrput_med_min_max_zoom}
    \vspace{-5pt}
\end{figure}

% \begin{figure}[!t]
% \centering
% \begin{subfigure}[t]{\columnwidth}
% \centering
% \includegraphics[width=\columnwidth,keepaspectratio=true]{images/22a.pdf}
% \phantomsubcaption
% \end{subfigure}
% \begin{subfigure}[t]{\columnwidth}
% \centering
% \includegraphics[width=\columnwidth,keepaspectratio=true]{images/22b.pdf}
% \phantomsubcaption
% \end{subfigure}
% \caption{Blow-up of the DMA MM2S throughput results from the left-hand side of Fig. \ref{fig:DMA_S2MM_thrput_med_min_max} emphasizing behavior for smaller payload sizes.}
% \label{fig:DMA_S2MM_thrput_med_min_max_zoom}
% \vspace{-0pt}
% \end{figure}

Figs. \ref{fig:DMA_MM2S_latency_med_min_max}, \ref{fig:DMA_MM2S_thrput_med_min_max} and \ref{fig:DMA_S2MM_thrput_med_min_max} plot the median-min-max results for MM2S latency, and for MM2S and S2MM throughput, respectively, using data from 10,000 trials. Experimental results for payloads of size 4 B through 1 MB are superimposed. The median values are plotted as curves through shaded regions delineated by the measured minimum and maximum values. Note that the median and maximum curves are often coincident (and are indistinguishable), which indicates that the occurrence of minimum throughput is a rare event. Figs. \ref{fig:DMA_MM2S_thrput_med_min_max_zoom} and \ref{fig:DMA_S2MM_thrput_med_min_max_zoom} blow-up the region for payloads between 4 and 128 B to better portray throughput for smaller payload sizes.

The MM2S latency results shown in Fig. \ref{fig:DMA_MM2S_latency_med_min_max} for the ZCU102 and ZCU111 are very similar, with the median and minimum latencies for the ZCU111 only slightly smaller than the ZCU102. As indicated earlier, the MM2S latencies include overhead associated with the execution of RPU C code, whereas the S2MM latencies do not. Although the S2MM latencies are not shown, they can be computed from the throughputs given in Fig. \ref{fig:DMA_S2MM_thrput_med_min_max_zoom} using the smallest payloads of 4, 8 and 32 B, corresponding to the DMA data-bus bit-widths, respectively. The RPU C code execution overhead is significant with MM2S latencies measured at 1.3\mics\!, whereas the S2MM latencies are smaller by nearly a factor of 10 at 136 ns.

The MM2S throughput results shown in Figs. \ref{fig:DMA_MM2S_thrput_med_min_max} and \ref{fig:DMA_S2MM_thrput_med_min_max} show similar one-sided performance metrics, with nearly coincident median and maximum throughput values and distinct minimum throughput values. The benefits of the larger and faster DDR within the ZCU111 are most apparent for the largest DMA data-bus bit-widths and payload sizes. For example, the maximum ZCU102 MM2S throughput is 4.5 GB/s for bit-widths of 256 and 1024, whereas the maximum for the ZCU111 increases to 10.5 GB/s and 19.2 GB/s, respectively. The S2MM results are similar except the throughput for the ZCU111 is maximum at 17.1 MB/s with the DMA engine configured with a bit-width of 1024, and it exhibits larger variability. Interestingly, the variation in the throughput rates approaches 0 for the largest payload sizes for any DMA bit-width, which can be leveraged when TIQC systems require continuous raw gate sequence reconfigurations.

\subsection{\change{CDMA}: PS-DDR to PL-DDR}
\begin{figure}
    \centering
    \includegraphics[width=\columnwidth,keepaspectratio=true]{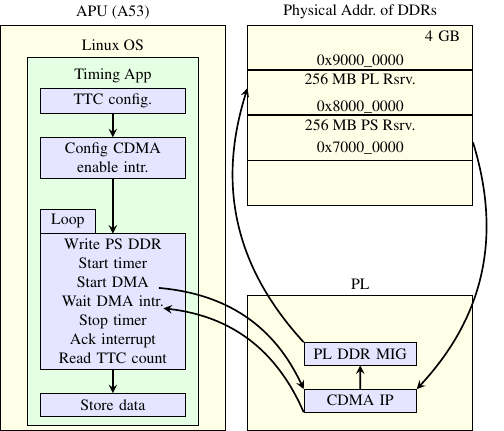}
    \caption{Flow diagram for CDMA between the PS and PL DDRs.}
    \label{fig:CDMA_PS_to_PL_DDR_block_diagram}
    \vspace{-5pt}
\end{figure}

A second type of DMA operation investigated in this paper is referred to as central DMA (CDMA), and is annotated with \tcirc{4} and \tcirc{5} in Fig. \ref{fig:Zynq_communication_channels}). CDMA handles block-level data transfers between PS and PL DDR memories. The flow diagram of the test procedure is shown in Fig. \ref{fig:CDMA_PS_to_PL_DDR_block_diagram}. The APU first configures the \textit{Triple Timer Counter} (TTC) and CDMA engine with the PS DDR source and PL DDR destination addresses. The Loop component carries out multiple repeated trials of the DMA transfer operation. The first component of the loop writes random values into the PS DDR memory region (assigned a physical address of 0x7000\_0000 within the memory map shown on the left). The timing interval is annotated by the 'Start timer' and 'Stop timer' labels in the figure. The CDMA transfer operation is sandwiched between these statements which is initiated by writing the length register within the CDMA controller. The CDMA engine generates an interrupt to indicate that the transfer has completed. 

Note that the Linux kernel requires a specialized device-tree configuration with reserved memory sections for both the PS and PL DDR memories to prevent Linux from utilizing these DMA source and destination regions as part of its virtual memory system. The CDMA engine itself is configured as an IP block in the PL of the ZCUs, and possesses the same set of configuration parameters as the DMA engine discussed earlier, i.e. input system clock frequency and DMA data-bus bit-width. Similar to the DMA experiments described in the previous section, we created a set of bitstreams with different configurations. In particular, 32, 64 and 256 bit-width versions are created, each with the system clock frequency set to 300 MHz and 333 MHz for the ZCU102 and ZCU111, respectively.

\begin{figure}
    \centering
    \includegraphics[width=\columnwidth,keepaspectratio=true]{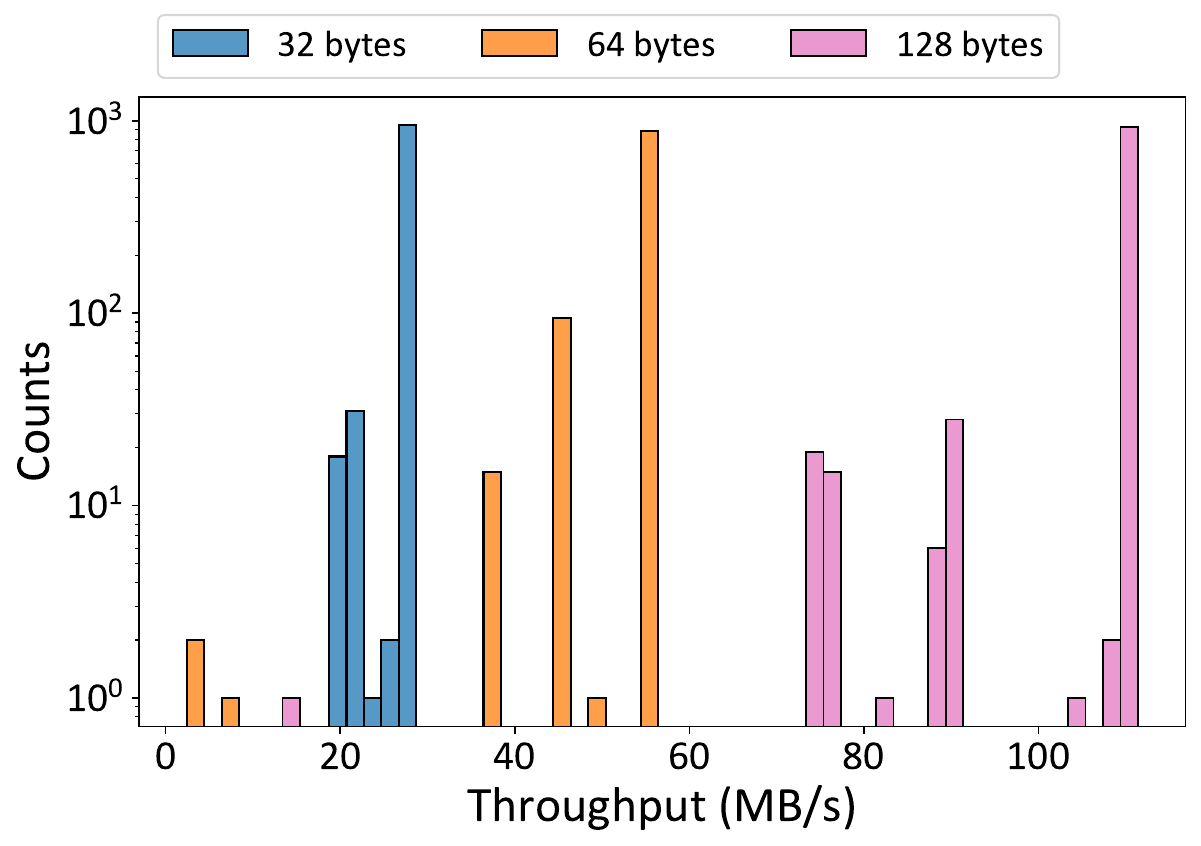}
    \caption{Histogram showing results for CDMA running on the ZCU111 with the PL clock frequency set to 333 MHz, DMA bit-width set to 256, and for payload sizes of 32, 64 and 128 B. }
    \label{fig:CDMA_PS2PL_thrput_raw}
    \vspace{-15pt}
\end{figure}

\begin{figure}
    \centering
    \includegraphics[width=\columnwidth,keepaspectratio=true]{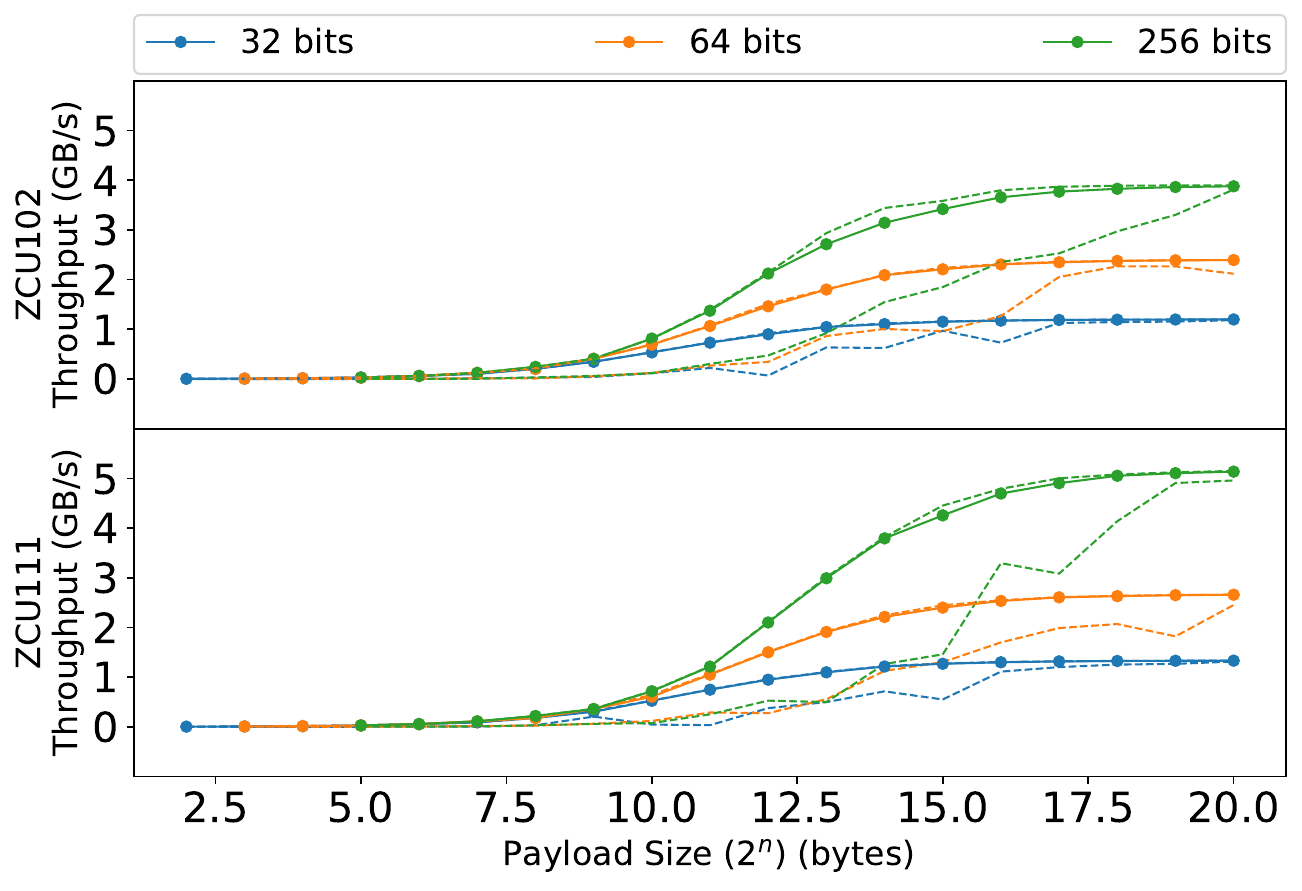}
    \caption{CDMA throughput results for the ZCU111 PL running at 333 MHz and the ZCU102 PL running at 300 MHz for payload sizes from 4 B to 1 MB and for DMA bit-widths of 32, 64 and 256.}
    \label{fig:CDMA_PS_to_PL_DDR_thrput_med_min_max}
    \vspace{0pt}
\end{figure}

The multi-user, multi-tasking nature of the Linux OS adds variability to the measurements, when compared with the RPU, as expected, and the maximum overall throughput is lower. A histogram showing the throughput results derived from data collected from 1000 trials, and for small payload sizes of 32, 64 and 128 B, is shown in Fig. \ref{fig:CDMA_PS2PL_thrput_raw}. Although most of the minimum throughputs occur as a fraction of 25\% or less of the median and maximum values, several trials show significant deviations. The minimum throughputs over an extended run of 100,000 trials, and for payload sizes from 32 B to 1 MB are shown in Fig. \ref{fig:CDMA_PS_to_PL_DDR_thrput_med_min_max}. Although the root cause of the slowdowns is attributable to interrupt service routine calls within the Linux kernel, which occur between the sequence of operations carried out during the timing operation, such behavior is unavoidable within Linux OS environment, unless all interrupts are disabled during this call sequence or Linux is replaced with a bare-metal application. The latter solution will reduce the variability to values similar to those shown for the RPU (see Fig. \ref{fig:DMA_MM2S_thrput_raw}), but it will also eliminate convenient access to system services provided by Linux to user applications.

% \begin{figure}[!t]
% \centering
% \begin{subfigure}[t]{\columnwidth}
% \centering
% \includegraphics[width=\columnwidth,keepaspectratio=true]{images/25a.pdf}
% \phantomsubcaption
% \end{subfigure}
% \begin{subfigure}[t]{\columnwidth}
% \centering
% \includegraphics[width=\columnwidth,keepaspectratio=true]{images/25b.pdf}
% \phantomsubcaption
% \end{subfigure}
% \caption{CDMA throughput results for the ZCU111 PL running at 333 MHz and the ZCU102 PL running at 300 MHz for payload sizes from 4 B to 1 MB and for DMA bit-widths of 32, 64 and 256.}
% \label{fig:CDMA_PS_to_PL_DDR_thrput_med_min_max}
% \vspace{-0pt}
% \end{figure}
\section{Analysis of performance limitations}
\label{sec:limits}

TIQC control system requirements for communication rates in the DMA core depend on multiple factors, including the number of bits used to define parameters, the duration and complexity of pulses, the frequency at which they are updated to reflect calibration measurements, and how often intra-algorithm measurements affect future gates. Moreover, the gate sequences themselves can be represented in raw form and in a compressed format. Here we show that communication rates measured in section \ref{sec:com} are sufficient for correcting worst case scenarios where all qubits must be modified simultaneously.

We start by evaluating a normal operating scenario where gate parameters are preset.  The compressed gate representation leverages a principle in computer architecture called temporal locality, where a series of LUTs are pre-loaded with data from gate sequences that are likely to be reused in the near future. The concept is also described in the appendix (\ref{sec:gate_sequencers}) in reference to Fig. \ref{fig:luts}. In the compressed representation, a gate is represented by an 11-bit identifier (ID), and up to 20 gate IDs can be packed into a 256-bit word (with the remaining 36-bits used for metadata). 

The bandwidth requirement for streaming predetermined gate sequences is therefore reduced by a factor of 160, from $\theta_{DMA} \equiv 10.656~\text{GB/s}$ (see below) to $\approx 66.6~\text{MB/s}$. 
Though the throughput reduction is significant, it still exceeds the maximum throughput available for AXI-Lite GPIO, which was specified earlier to be 41.9 MB/s in reference to Fig. \ref{fig:GPIO_RPU_thrput_config2}. DMA, however, is sufficient to meet the bandwidth requirements for both the raw and compressed gate sequence representations.  
%\change{There are however limits on how short the gates or segments of gates can be, the latter being applicable in a case like Trotterized gates, where back-to-back pulses that comprise a gate cannot be any shorter than 19.5~\text{ns}, limited by the gate sequencer.}

A more demanding scenario occurs where a gate parameter must be changed simultaneously for all channels prior to the next gate, \change{for instance when a laser that is used for multiple ions suddenly drops in power} and the modulation signal sent to all affected AOMs must compensate \change{by changing the amplitude spline parameters. Another example is when a prior measurement (like error correction or drift control) requires updating subsequent gate parameters.} There are 
%algorithmic
other approaches to dealing with such rare events, but for simplicity we place the burden on the pulse generation part of the control system because it can apply tailored corrections on a per-qubit basis. Although some pauses are acceptable, we seek to achieve a response that is on the order of the fastest gate time, assumed here to be 1\mics\!.  \change{While this time is much shorter than currently achieved in typical experiments, we use it to analyze the suitability of this control system for larger scale TIQCs, where reducing latency will be critical.}  Based on the direct streaming mode described in section \ref{sec:dds}, gates use a minimum of 8 parameters per channel, yielding a total of 64 parameters that need to be executed in parallel at any given time. Each parameter is represented as a 256-bit word that encodes spline coefficient data, regardless of whether the parameter is modulated or constant. Assuming the streaming input side of the FIFO is clocked at 333 MHz (see Fig. \ref{fig:GateSequenceBlockDiagram}), this requires a throughput of $\theta_{D\!M\!A} \equiv W_{bus} f_{D\!M\!A} = 10.656~\text{GB/s}$. Alternatively, we can instead cast this into a gate throughput, $\theta_{G}$, where the effective data size is $W_G = 64 W_{bus}$ and $\theta_G = \theta_{D\!M\!A}/W_G = 5.203 \times10^6~\text{gates/s}$. The shortest gate time which can be continuously streamed is thus $1/\theta_G = 192.2$~\text{ns}, neglecting the time required to compute an update to the parameters (which could be longer than 1\mics\!).

For the compressed gate mode (using the GLUT described in the appendix \ref{sec:appendix_dds}), an individual gate can be programmed and sequenced on a single channel with a minimum of 11 words, 8 words for the pulse LUT, 1 word for the memory map LUT, 1 word for the gate LUT, and an additional word for reading out the gate. In this case, the number of words, $N_w = 11$, needed for all channels, $N_{ch}=8$, leads to $W_G = N_{ch} N_{w} W_{bus} = 88 W_{bus}$, and $1/\theta_G = 264.3~\text{ns}$. \change{This is particularly relevant for cases where all parameters are updated but then remain constant for many subsequent gates, e.g. when there is a slow drift in laser power.}

Once the initial programming data are sent, the gate can be read out with a single word per channel. Supposing all parameters of the gate need to be updated on each execution of the gate, the subsequent gate calls can be made with 9 words per channel by modifying the data in the pulse LUT prior to reading out the gate. In most cases, the modified gate data will be restricted to a single, or perhaps a small subset of channels. 
Instead, it is useful to think of the number of sequencing words needed per channel, S, and the total number of parameters that need to be updated across all channels, $P_{upd}$. The total number of words that needs to be transferred on each iteration is then $N_w = S N_{ch} + P_{upd}$. For back-to-back execution, with a full update of an 8-parameter gate, we have $N_w = 16$ and a minimum gate time of $\theta^{-1}_G = 48~\text{ns}$. If only a single parameter is updated for a single channel, this yields $N_w = 9$ and a minimum gate time of $\theta^{-1}_G = 27~\text{ns}$.

However, taking advantage of the ability to pack multiple gate identifiers in a single transfer can reduce the minimum gate time to $\theta^{-1}_G = 28.1~\text{ns}$ when updating 8 parameters, and $\theta^{-1}_G = 7.1~\text{ns}$ when updating a single parameter. \change{The direct streaming rate (meaning all parameters are preset and not updated) is 1.2~ns, however this is not achievable because the minimum gate time for continuous operation is limited by the gate sequencer to 19.5~ns.  As an example this limit applies to the situation where pre-determined Trottererized segments of a gate are reduced to 19.5~ns and locally stored in the LUTs.}

An alternative limit can be imposed by calculating the number of parameters which can be updated when running 1\mics gates. For updates interleaved between each gate, the maximum number of parameters that can be modified is $f_{D\!M\!A}(1 \mu\text{s})-8 = 325$. Similarly, if one wants to update all 32,768 values in the pulse LUTs, this can be achieved if the programming data is run after 99 sequential 1\mics gates.  \change{An example where this is relevant is when many parameters have to be updated but not necessarily right away.}

These limits assume a sufficiently large payload size such that the programming and sequencing data are densely packed, as well as pre-determined albeit potentially changing parameters. However, if there are a small number of parameters which are not known in advance and need to be updated before the next sequence, then this may require a small DMA transfer. For a single parameter, encoded in 32 bytes, latency dominates the overall throughput, where the minimum transfer rate for a 32-byte payload size is 17.6 MB/s, as shown in Fig.~\ref{fig:DMA_MM2S_thrput_med_min_max_zoom}. This corresponds to 1.82 \mics for the fastest time to update a single parameter, again neglecting calculation times. More parameters could be transferred in roughly the same time by using a larger payload size.  

This FIFO clock speed is a maximum rate, so measuring outlier slow communication rates is important for determining the realistic limitations of the system.  The curves from Fig. \ref{fig:DMA_MM2S_thrput_med_min_max} show that the median (and minimum) DMA transfer throughput of the ZCU111 with the DMA IP block configured with a 256-bit width and run at 333 MHz is $\approx 10.5~\text{GB/s}$, which is slightly less than $\theta_{D\!M\!A}$. 
However, with the DMA IP block configured with a larger bit width of 1024, the median (and minimum) throughput is $\geq 19.0~\text{GB/s}$, which supports the maximum throughput with additional headroom.
Given that the typical fastest gate time for a TIQC is about 1\mics (neglecting the short times used for virtual $Z$ gates), even the 256-bit DMA width would be sufficient.

For the extreme case, where gate sequences reference pulse information that is completely unique and gate information cannot be reused, one may consider using a direct streaming mode in which the gate sequencer LUTs, discussed earlier in reference to Fig. \ref{fig:luts}, are bypassed. In the limit of the shortest possible gates that can be continuously streamed, this approach may be preferable to constant reprogramming of the LUTs. Although the LUTs can be programmed in a way that effectively treats the gate sequencer as a deep FIFO, increased FIFO depth is immaterial in situations where the feed rate matches the consumption rate of the spline engines, and the additional programming and sequencing data cut down on the maximum effective throughput.  This scheme is less flexible at correcting gate parameters, for instance the amplitude, and would instead require a full recalculation of all points.

\change{Recalculation is typically expensive, owing to the fact that spline coefficients need to be refitted for continuous modulations (especially when accounting for unavoidable non-linearities in AOM and amplifier response). Operating in a regime where large amounts of unique gate data need to be regularly regenerated will likely lead to bottlenecks at the APU, assuming gate data can be recalculated on chip, or possibly limited by network transfer if the calculations demand the computing power of an external server. In these cases, throughput is dominated by classical algorithmic efficiency, and potentially network throughput, both of which suffer from larger variability in timing.}

\change{However, this problem is offset by performing gate calculations while quantum circuits are running, and maintaining efficient compressed representations of gates, in which changes to one parameter will affect multiple gates. Partial reprogramming of LUTs allows the RPU and PL to coordinate quantum circuits with classical control flow while the APU is free to generate the next set of gate data, thus maximizing the benefits of AMP. Because the APU can in most cases fully recalculate compressed gate data faster than the duration of a two-qubit gate ($\approx 200~\mu\text{s}$)~\cite{lobser:2022}, the remaining transfer overhead is less of a dominating factor in this mode of operation.}

\change{The measured transfer latencies are several orders of magnitude smaller than coherence times in TIQC systems, particularly for $^{171}\text{Yb}$ where coherence times typically range anywhere from 1-1000 s, giving a lot of headroom for classical control flow from the RPU that depends on mid-circuit measurements. For systems that are relatively stable and circuits that can leverage redundancy in gate data, the measured timing characteristics are well within typical requirements for gate throughput for most applications.}

% \change{As described in the preceding paragraphs, there are many conditions and caveats that affect the speed of the control system.  We summarize several important ones in Table \ref{tab:summaryTable} for clarity.}

% \begin{table}
%     %\centering
%     \begin{tabular}{ | p{0.2\columnwidth} | p{0.3\columnwidth} | p{0.3\columnwidth}| }
%         \hline\hline
%         \textbf{Condition} & \textbf{Speed limit} & \textbf{Explanation} \\ 
%         \hline
%         Condition & Time limit & Explanation \\ 
%         \hline
%         Condition & Time limit & Explanation \\ 
%         \hline\hline
%     \end{tabular}
%     \caption{\raggedright \change{write a caption}}
%     \label{tab:summaryTable}
% \end{table}
\section{Conclusion}
\label{sec:conclusion}

We draw the following conclusions regarding the applicable usage scenarios and limits of the communication mechanisms of a Xilinx Zynq MPSoC and RFSoC within the context of a TIQC system architecture:
\begin{itemize}
    \item RPU-driven DMA requires a PL clock frequency of 333 MHz and a width greater than 256 bits, e.g. 512 or 1024 bits, to meet the 10.656 GB/s requirement for streaming raw, uncompressed, gate sequences.
    \item RPU-driven DMA must be used for raw or compressed gate sequences but can be relaxed for the latter case, e.g. by using a DMA width of 256-bits and/or lower PL-side clock frequencies.
    \item RPU-driven GPIO can be used to meet soft real-time, lower-bandwidth system requirements for qubit components running in the PL. For example, shuttling, Doppler cooling, state detection, or ion reloading.
    \item APU-to-RPU RPMsg throughput is higher and exhibits lower variability than the corresponding metrics computed for RPU-to-APU transfers, but nonetheless can only be used to meet low bandwidth soft deadline-based requirements. For example, we intend to use RPMsg for the transfer of control and status information between the APU and RPU, and for computing shift deltas for control parameters via a feedback algorithm.
    \item APU-driven CDMA transfers between PS and PL DDRs exhibit high variability under the Linux OS. However, minimum throughputs approach 5.0 GB/s for large payload sizes (> 1 MB), which enables APU updates to compressed gate sequences to be transferred to PL DDR with plenty of headroom to meet data consumption rates for RPU DMA transfers to PL. Note that the APU will host the pulse compiler for generating compressed sequence data that needs to be transferred to PL DDR for access by the RPU.
    \item A hopefully rare worst-case scenario that nonetheless should be accommodated by the control system occurs when one or more gate parameters need to be updated on the fastest timescale of the quantum computer.  Limited by latency, we find that this architecture can update gates in less than 2\mics\!.
\end{itemize}

Even with relatively long gate times, the electronic control system for a TIQC must be designed with communication throughput in mind in order to achieve near real-time updates on gate parameters. The design and measurements described in this paper are specific to the MPSoC and RFSoC used here but can be translated to similar hardware to identify limits on full channel updates and other performance scenarios for other qubit technologies.  Although the architecture we describe is most applicable for large scale quantum computing, understanding the hardware limitations of electronic control systems and testing them on current NISQ systems \cite{preskill:2018} will motivate theoretical, experimental, and engineering research to overcome them.
\section{Appendix}
\label{sec:appendix}

\subsection{DDS design}
\label{sec:appendix_dds}
The custom DDS core can generate two RF tones in order to drive Raman transitions and bichromatic two-qubit M\o lmer-S\o rensen gates that are commonly used in TIQC.
These tones are added in the digital domain to sidestep frequency-dependent phase shifts and amplitude distortion effects inherent to external RF components (i.e. combiners and mixers).
Inputs include frequency, phase, and amplitude words for each tone, as well as a number auxiliary inputs, such as single-bit inputs for triggering phase synchronization or enabling feedforward corrections.
Both DDS tones are set up in an interleaved configuration to double the effective sampling rate while maintaining an input frequency of 409.6 MHz, which is below the maximum AXIS clock speed of 500 MHz and makes use of the RF data converter (RFDC) core's 8$\times$ interpolation filter for using the maximum sampling rate of 6.5536 GSPS for the ZCU111 DAC outputs\footnote{Maximizing the sampling rate allows for digital up-conversion (DUC) of the input frequencies with the RFDC's numerically-controlled oscillators (NCOs) to provide the largest allowable range of baseband frequencies.}.

The main distinction between the custom DDS design and a conventional DDS design is the inclusion of three specialized features: global phase synchronization, frequency feedforward corrections, and elements used to compensate for cross-talk errors at the experiment level.

\subsubsection{Global phase synchronization}
Global phase synchronization is a feature that allows reuse of DDS cores for driving different frequencies, and the ability to return to a previous frequency and phase as if the DDS had been in a free-running state.
Although this can be performed by calculating the expected phase and either overwriting the accumulator or adding a phase offset, the distinction here is that the global phase synchronization is handled automatically.
This removes the need for manual bookkeeping and also avoids any potential issues that may arise in the event of a missed clock edge or non-deterministic latency.
Eliminating manual bookkeeping requirements also leads to a smaller data footprint, since gates can be represented as simple primitives that can be reused without having to account for context-dependency.
The process involves a global counter that is shared among all DDS cores, and the counter data is multiplied against the DDS input frequency to calculate the phase accumulated from some arbitrary point in the past when the global counter is zero.
The resulting phase is passed to the DDS accumulator, with latencies matched in the data path such that a trigger can update the accumulator with the global phase corresponding to the current input frequency.
This allows one to synchronize to the global phase across all channels, at any point in time, and the resulting phase will be consistent with a previously synchronized frequency of the same value.
Reproducing the phase simply boils down to an initial synchronization step with the first application of some given frequency in a gate sequence.
This is possible without ever flushing out the accumulator, so that a frequency can be synchronized to its previous phase for as long as the device is powered and the global counter is not reset\footnote{This is possible even if the global counter rolls over because of commensurability over a finite set of frequency values limited to $2^N-1$ for an $N$-bit frequency word.}.

\subsubsection{Frequency feed-forward corrections}
Drift in the frequency or repetition rate of the gate laser is a common error that can be corrected by the electronic control system.  For example, due to drift in the cavity length of the pulsed laser used for ytterbium qubit operations, the repetition rate is actively monitored to account for ``breathing'' in the spectrum from the frequency comb.
To track the resulting frequency shift and accumulated phase in the repetition frequency, we use a scheme similar to~\cite{mount:2016}. The overall error between two harmonics in the frequency comb can be accounted for by adding an offset to the DDS accumulator output, which is read out of the accumulator of a DDS in a dedicated frequency feedback module, and subsequently multiplied by the harmonic separation in the comb that is nearest to the target transition frequency.

\begin{figure*}[htbp]
  \centering
  \includegraphics[width=6.0in,keepaspectratio=true]{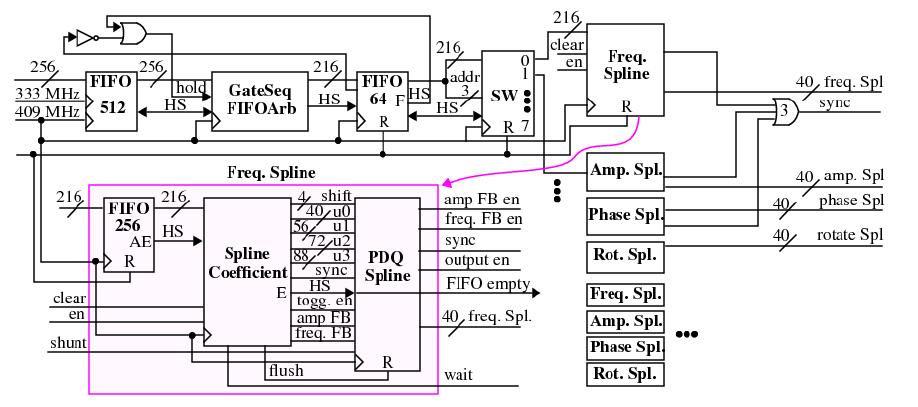}%{images/Gate_sequencer_block_diagram.ps.pdf}
  \caption{Gate sequencer block diagram. The DMA-based streaming AXI interface is connected to the upper, left-most 256-bit FIFO input in the diagram. The cubic spline lightweight interpolation engines for the 8 parameters associated with the two independent tones of the DDS core are shown along the left side, and expanded for the frequency spline within the magenta rectangle.}
  \label{fig:GateSequenceBlockDiagram}
\end{figure*}

\subsubsection{Crosstalk compensation}
Crosstalk-induced errors resulting from optical overlap of individual addressing beams with nearest- and next-nearest-neighbor qubits can be accounted for with a cancellation tone that destructively interferes with the light from a neighboring beam. Although some schemes (like those that use a chain of ions in a single well \cite{clark:2021}) are more sensitive to this mechanism of crosstalk, similar effects exist for other trapped ion schemes and qubit technologies.
Additional crosstalk errors can arise from electrical crosstalk driving the AOM transducers as well as sympathetic vibrations across crystals in a multi-channel AOM.
Electrical and acoustic crosstalk require a coarse delay adjustment to approximate the longer propagation time.
Fine-tuning the delay is approximated by adjusting the phase, which is achieved by using a complex-valued scaling of the input signals from neighboring channels to give an overall change in amplitude and phase.
To ensure that optical crosstalk errors are accounted for, the arrival times of crosstalk signals must be perfectly aligned with the output of the source DDS.
\change{This is accomplished by sharing each channel's ideal signal (meaning the codeword description of the intended signal without crosstalk) with each of its neighbors.  Cancellation tones are generated by multiplying the shared signals by a complex factor that shifts the phase by the desired amount (typically $\pi$ plus smaller perturbations due to alignment imperfections) and attenuates it to account for the pickup ratio.  Because these calculations take time, a delay is added to the offending channel so as to align the cancellation tones with its ideal signal.  The cancellation tones are added to their own ideal signals.  Since every channel can be an offending channel, all channels delay their actual output.}
% To accomplish this, a secondary signal is output from the DDS, which is passed to the neighboring DDSs, and the primary signal passed to the RFDC is run through a delay line with a latency that matches the DSP elements used for applying the complex phase shifts.
% The phase-correct signal is then added to the primary output of the DDS.

\subsection{Gate sequencers}\label{sec:gate_sequencers}
To reduce the amount of information needed to express complex modulations, the data is cast into cubic spline coefficients which are interpolated by lightweight spline engines in the gate sequencers \cite{bowler:2013}.
Spline coefficients are independently specified and interpolated for each waveform parameter, where each segment is encoded in 216-bit words with the form
\begin{equation*}
\left\lbrace M, \tau, U_3, U_2, U_1, U_0\right\rbrace ,
\end{equation*}
where $U_n$ are the spline coefficients, $\tau$ is the the duration (or number of clock cycles) to interpolate, and $M$ is additional metadata.
In order to maintain concurrent operation of all spline engines, the segments are buffered using FIFOs, which are fed on timescales of the system clock over the number of parameters, $T_{clk}/N_p$, and consumed on timescales given by $T_{clk}/\tau$.
The values for $\tau$ can vary depending on gate durations and the number of knots used to specify the modulation, but are often on the order of 10-100, which is larger than the number of parameters $N_p = 8$.
Once the FIFOs have been populated, spline engines are enabled via a global trigger to ensure concurrent operation. A block diagram of the gate sequencer pipelined architecture is shown in Fig. \ref{fig:GateSequenceBlockDiagram}.

The 8 parameters are frequency, phase, amplitude, and a ``frame rotation'', specified for two independent tones in the DDS core.
Frame rotations are used to represent a third degree of freedom which is not directly accessible from an individual addressing beam.
Since $\hat{x}$ and $\hat{y}$ dimensions in the qubit's Bloch sphere \cite{nielsen:2010} can be accessed via a change in laser phase, it is possible to virtualize $\hat{z}$ rotations by shifting the phase of subsequent gates in a circuit.
Although this can be done by pre-calculating the effective gate sequence when $Z$ gates are present, this poses challenges for circuits that use mid-circuit measurements where $Z$ gates are conditionally applied.
By abstracting these $Z$ operations with a parameter that tracks the desired frame of the qubit, we eliminate the need for manual bookkeeping and reduce the amount of unique gate information needed to encode a long circuit. Frame rotations are thus implemented as a cumulative phase, where the inputs are applied normally and added to an accumulator at the end of a pulse \change{so that it is treated on the same footing as the conventional phase, e.g. $\exp(i (\omega t + \phi + \phi_{frame}))$}\footnote{Appending at the end of the pulsed allows for special cases such as adjusting the qubit frame associated with AC Stark shifts during an ORE-correcting pulse such that the qubit frame is only adjusted for subsequent gates \cite{lobser:2022}. Additionally, splines can be used in which only the final value should be accumulated. The use of spline-modulated frame rotations is quite useful in that they can be calculated from the integral of the amplitude modulation and scaled (or for non-linear effects, conformally mapped) to track the AC Stark shift during a pulse. This can greatly simplify global phase synchronization, accounting for amplitude-dependent frequency shifts, and calibration since the scale factor can be determined from the overall phase shift accrued by the qubit.}.

To maintain consistency in the firmware design, the data path is made as uniform as possible for all gates and parameters.
This means that simple square-pulse gates, or any parameter that is constant for the duration of the gate, is represented by data for which the higher order spline coefficients are set to zero.
Gates are often repeated multiple times throughout a circuit; by abstracting away frame rotations, as well as automating global phase synchronization, they can be described with a single representation that is devoid of any context dependency.
Additionally, most gates, especially single-qubit gates, have nearly identical representations, where differences in rotation axis only affect an overall phase offset.
This leads to a large amount of data redundancy, which can be locally stored in lookup tables (LUTs) and read out using a more compact representation.

The LUTs are set up in three separate stages to minimize the data needed to stream out a fast gate sequence. The hierarchy and connectivity relationship among the three LUT types is shown in Fig. \ref{fig:luts}.
The lowest level ``Pulse LUT'' (PLUT) stores the 216-bit spline segment data, which is distilled down to unique segments shared across all gates on that output channel.
Because the data in the PLUT is unique, the ordering of data is completely arbitrary.
This arbitrary ordering is reconciled using a second ``Memory Map LUT'' (MLUT), which represents non-contiguous and repeated entries of the PLUT by storing PLUT addresses in a linearly-ordered address space of the MLUT.
The MLUT allows gates to be represented as a pair of start and end addresses that can be stepped through sequentially.

One more layer of compression is used to store gates in a ``Gate LUT'' (GLUT), where gates are given a unique GLUT address, and the resulting data is a concatenated word containing the start and end addresses in the MLUT.
Gate identifiers are densely packed into single 256-bit input words, which contain additional metadata for routing, the number of gates contained in the word, and data which indicates that the word contains gates for reading out of the LUTs.
The word are consumed in 11-bit segments, and passed to the GLUT.
The output of the GLUT is passed into an iterator module that steps through the start and end addresses.
Addresses from each iteration are passed to the MLUT, whose output is connected directly to the PLUT.
Raw segment data coming out of the PLUT is then routed to the appropriate spline engine FIFOs using the segments' metadata.

Because circuits can require large numbers of gates (on the order of $10^4$ for noisy intermediate-scale quantum (NISQ)  devices and many orders of magnitude more for demanding simulations \cite{lee:2021}), this type of compressed representation offers significant gains in data throughput.
Even the most simple of gate representations requires 2 kbit
%kib 
of data to feed all the FIFOs and maintain constant FIFO filling for concurrent operation. However, reducing the representation of this gate to 11 bits only requires two extra programming words (in which MLUT and GLUT data can be packed into single 256-bit transfers) and a streaming word.
The up front cost for encoding a single gate is immediately accounted for with the second application of the gate, as well as in cases where most of the gate data is shared.
Moreover, a large portion of circuits are run repeatedly to accumulate statistics on the measurement outcomes, particularly for calibration routines which comprise the majority of experimental runs on NISQ devices.
The compression scheme used by the LUTs has a clear advantage for reducing bandwidth requirements for sequencing large numbers of circuits with tens to hundreds of averages.
These gains are twofold when accounting for memory representations of the sequence data alone, which is imperative for successful interoperability between the PL, APU, and RPU when running classically-conditioned sequences.

The gate sequencer LUTs are essential for generating a seamless architecture that supports deterministic timing and hybrid algorithms.
Compiling the compressed gate sequence data creates a trade-off between data size reduction and classical algorithmic complexity.
Performing the compilation on an external server adds latency to the experimental cycle.
This can have greater impact on high-level feedback algorithms that execute classically-conditioned sequences, or algorithms that actively mutate gate definitions to shim out slow drift in calibration parameters, a feature that becomes increasingly vital as the number of qubits, and thus the number of calibrated parameters, grows.
Offloading the pulse compiler used for generating compressed sequence data to the SoC can offer a tighter feedback loop for these types of high-level feedback.
This approach also comes with the benefit that the classical resources required for compilation are predominantly fixed to the number of RF channels on the board, creating a distributed architecture with scalability built in.
However, it requires a compiler that is fast enough that it can either outpace the average duration of a typical experimental sequence, or offer comparable performance to an external server when accounting for network latency.

The classical computing power and memory constraints of the APU and RPU become relevant in the context of the LUTs used to encode pulses.
Because the LUTs themselves are implemented using half of the available Ultra RAM (URAM) primitives on the device---the URAM blocks are 288 kib, making up 2.8125 MB
%MiB 
of the total on-chip storage when accounting for the 80 available blocks---the total storage size outweighs the 256 kB
%kiB 
space allotted to the RPU's tightly-coupled memories (TCMs).
This poses additional challenges since a complete software representation of the gate data is necessary in certain situations.
For example, gate sequences with large amounts of unique data can exceed the allotted memory in URAM, requiring dynamic reprogramming mid-circuit.
Another requirement is an abstracted software representation for gate data that can be used to correctly mutate definitions at the appropriate memory locations to minimize reprogramming time.
%\comment{All: standardize 'i' or no 'i' in the abbreviations}

Our approach employs the APU as a math coprocessor, with responsibilities that focus on compiling the compressed pulse representations and programming the gate sequencer LUTs.
The sequence data, and any partial reprogramming data, is specified by the APU and written to a memory-mapped regions of DDR RAM.
The RPU schedules certain operations used for qubit state preparation and measurement, other time-critical classical operations such as shuttling ions, and initiating DMA transfers to burst the relevant data to the gate sequencers for particular sub-circuits.
Any operations that require callbacks to the APU are communicated via RPMsg, and the APU can optionally pass the callback to an external server if no local definition for the call exists.
Because the APU can in principle break timing determinism, the APU provides appropriate handshaking signals to indicate that the new sequences are ready.
However, the APU latency can be offset by the state preparation stage of an experimental sequence, which typically runs on the order of 1 ms.

Pipelining recompilation results with a fixed delay will add latency to the feedback loop, but can be used to maintain experimental duty cycle.
This option is fairly natural, since single-shot calibration measurements can be interleaved with normal experimental circuits, effectively increasing the time between experiments and allowing the compiler to update parameters before the next single-shot calibration.
To this end, the single-shot calibration measurements will typically result in small but predictable deltas (shifts) in a control parameter and thus a variation in the resulting gate data.
Designing a feedback algorithm that precomputes the possible deltas will allow the APU to have the appropriate data on hand as it is needed and, as a parameter drifts, computing new deltas well before they are needed.

On the other hand, certain high-level algorithms, such as variational quantum eigensolvers (VQE) or quantum approximate optimization algorithms (QAOA), often require more powerful computing resources.
These algorithms are in most cases either impossible or unreasonable to run on chip, but may be desired despite the increased latency and lack of timing determinism between shots.
However, nearly all of these high-level feedback routines will guarantee deterministic timing between state preparation and measurement, since the algorithms mentioned rely on results from a complete measurement of all qubits, in which case coherence times are no longer a bottleneck.
The potential for high-level feedback within a given circuit may be possible, but only in cases where outcomes are precalculated to a reasonable depth or coherence times are sufficiently long.
Regardless, one can maintain determinism by providing sufficiently long timeouts in which the RPU may still be able to perform other tasks and prepare certain register values and verify that a response has been received before the full timeout elapses, subsequently resuming the algorithm.
This architecture features the benefits of the PL, RPU, APU, and external control computers, with flexible and optional trade-offs in latency.

\subsection{Toolchain for Performance Measurements}
\label{sec:appendix_toolchain}
In this section, we outline the process we follow to create the experimental designs (the toolchain) from which the latency and throughput measurements are made. The process for each design involves a diversified sequence of steps, from creating and configuring IP blocks in a block diagram and writing VHDL descriptions of state machines, to Linux kernel building and device-tree configuration, through application coding and compilation. Open source APIs including Libmetal and OpenAMP are utilized across the tool chain. 

\begin{figure*}
   \centering
   \includegraphics[width=\linewidth,keepaspectratio=true]{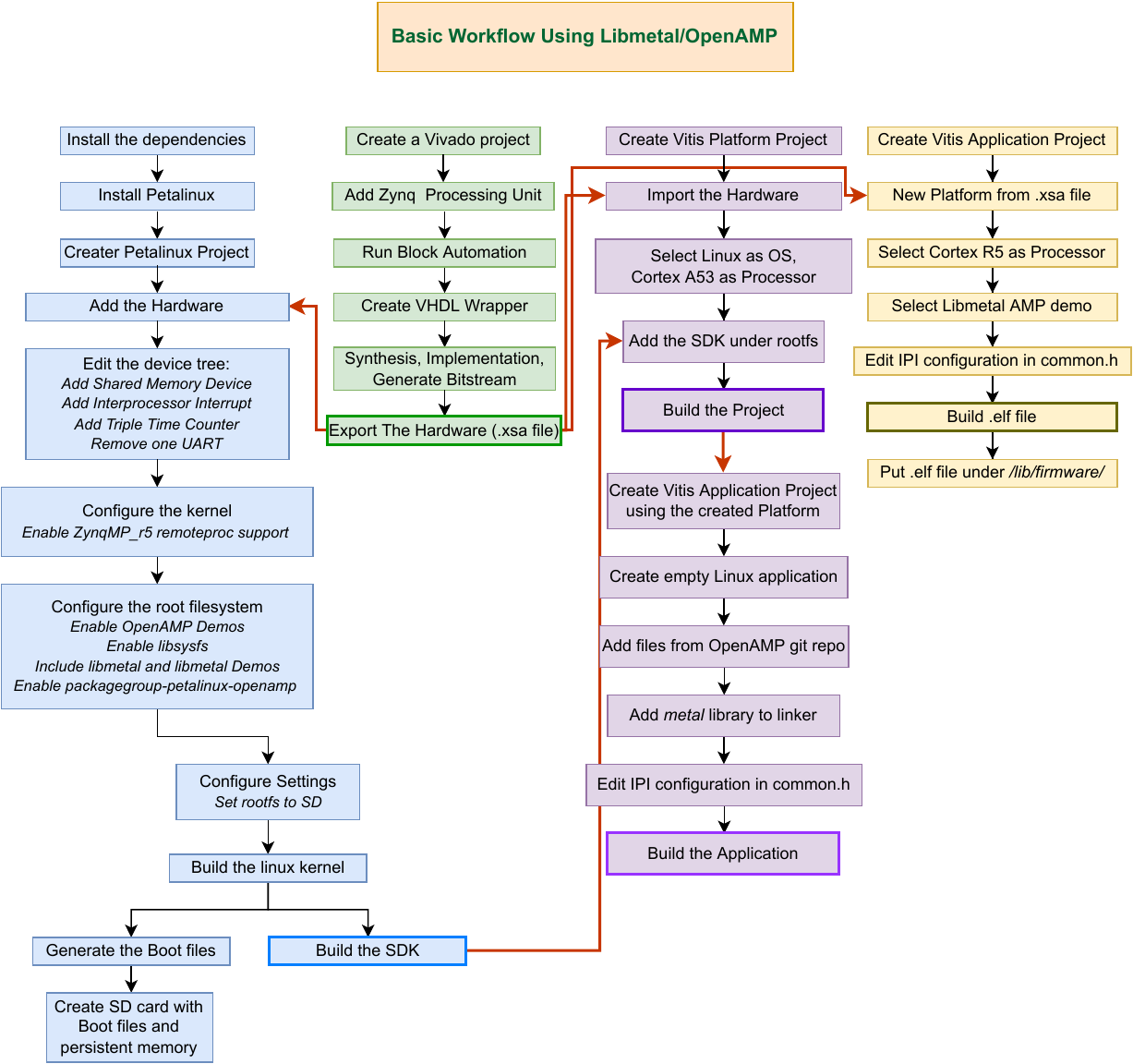}
   \caption{Linux kernel and OpenAMP application building process flow.}
   \label{fig:build}
\end{figure*}

We use the tool chain applications provided by Xilinx, including Vivado for synthesizing VHDL designs and for creating the block diagrams and programming bitstreams, Petalinux for configuring and building the embedded Linux kernel and device-tree components for the APU, and Vitis for creating the application binaries which run on the APU and RPU. A tool chain flow diagram is presented in \ref{fig:build}, where the steps for each of the Xilinx tool chain components are given in separate columns. In particular, the leftmost column shows the steps associated with using Petalinux, the next column illustrates the process flow used within the Vivado tool suite, whereas the last two columns show the steps followed within the Vitis embedded system application environment. Distinct process flows exist for building Linux application and bare metal applications.

Interdependencies between the tool chain components exist and are illustrated as red arrows in the diagram, where the output of a tool chain component is used as input to another. As shown in the figure, the hardware description file produced by the Vivado tool suite is used in the other tool flow components, and therefore defines a core component of any experimental design. Also, the Linux application design flow requires the kernel built by the Petalinux tools. The tool flow illustrated is not specific to our experimental designs, but rather represents a generic tool flow for any design and development board.

\subsection{Implementation Details for DMA: PL DDR to PL Streaming}

\begin{figure}
    \centering
    \includegraphics[width=3.05in,keepaspectratio=true]{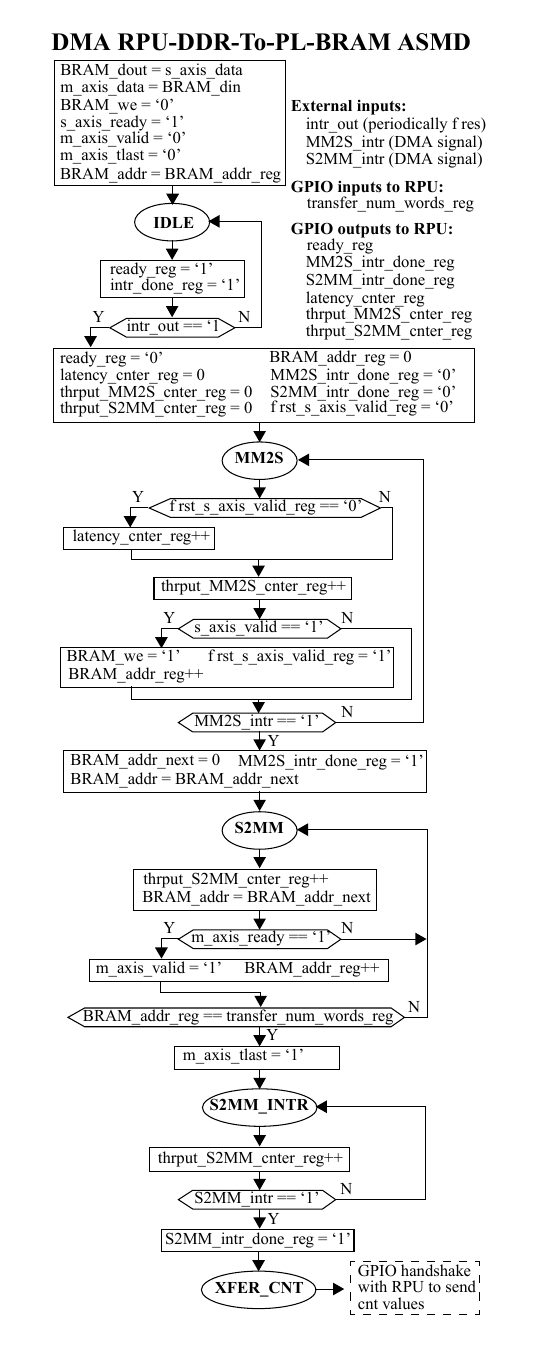}
    %{images/DMA_RPU_To_PL_Stream_ASMD.ps.pdf}
    \caption{DMA RPU-PL-DDR-to-PL-Stream ASMD.}
    \label{fig:DMA_RPU-PL-DDR-to-PL-Stream_ASMD}
    \vspace{-0pt}
\end{figure}

An algorithmic state machine diagram (ASMD) for the PL SM is given in Fig. \ref{fig:DMA_RPU-PL-DDR-to-PL-Stream_ASMD}. The AXI master and slave signals are controlled according to the rules of the AXI4 protocol, where assertions by the DMA engine of \textit{s\_axis\_valid} and \textit{m\_axis\_ready} are acknowledged in the same clock cycle by assertions of the PL SM AXI signals \textit{s\_axis\_ready} and \textit{m\_axis\_valid} (note that \textit{s\_axis\_ready} is held permanently at 1 to facilitate the maximum transfer rate). Data in and out of the PL BRAM takes place in one clock cycle, again facilitating the maximum transfer rate. The counters for throughput are stopped on reception of the DMA \textit{MM2S\_intr} and \textit{S2MM\_intr} signaling events.

\clearpage
\subsection{\change{List of Abbreviations}}
\protect\change{
\begin{itemize}
    \abbv{SoC}{System on a Chip}
    \abbv{DMA}{Direct Memory Access}
    \abbv{CDMA}{Central Direct Memory Access}
    \abbv{MPSoC}{MultiProcessor System On Chip}
    \abbv{RFSoC}{Radiofrequency System On Chip}
    \abbv{PL}{Programmable Logic}
    \abbv{PS}{Processing System}
    \begin{itemize}
        \abbv{APU}{Application Processing Unit}
        \abbv{RPU}{Real-time Processing Unit}   
    \end{itemize}
    \abbv{TIQC}{Trapped-ion Quantum Computer}
    \abbv{RF}{Radio Frequency}
    \abbv{AOM}{Acousto-Optic Modulator}
    \abbv{AWG}{Arbitrary Waveform Generators}
    \abbv{DDS}{Direct Digital Systhesizer}
    \abbv{FPGA}{Field Programmable Gate Array}
    \abbv{DAC}{Digital to Analog Converter}
    \abbv{ADC}{Analog to Digital Converter}
    \abbv{RFDC}{Radiofrequency Data Converter}
    \abbv{DSP}{Digital Signal Processing}
    \abbv{QEC}{Quantum Error Correction}
    \abbv{AXI}{Advanced eXtensible Interface}
    \abbv{ADAS}{Advanced Driver Assisted Systems}    
    \abbv{BRAM}{Block RAM}
    \abbv{OpenAMP}{Open Asymmetric Multiprocessing}
    \abbv{RPMsg}{Remote Processor Message}
    \abbv{RTOS}{Real-time Operating System}
    \abbv{GPIO}{General Purpose Input/Output}    
    \abbv{PLE}{Pulse Length Error}
    \abbv{ORE}{Off-resonant Error}
    \abbv{AM}{Amplitude Modulation}
    \abbv{FM}{Frequency Modulation}
    \abbv{GRAPE}{GRadient Ascent Pulse Engineering}
    \abbv{LUT}{Look-up Table}
    \begin{itemize}
        \abbv{PLUT}{Pulse LUT}
        \abbv{MLUT}{Memory-map LUT}
        \abbv{GLUT}{Gate LUT}
    \end{itemize}
    \abbv{IPC}{Inter-processor Communication}
    \abbv{IPI}{Inter-processor Interrupt}
    \abbv{TCM}{Tightly-Coupled Memory}
    \abbv{EMIO}{Extended Multiplexed I/O}
    \abbv{LCM}{Life Cycle Management}
    \abbv{SM}{State Machine}
    \abbv{FPD}{Full-Power Domain}
    \abbv{LPD}{Low-Power Domain}
    \abbv{TTC}{Triple Timer Counter}
    \abbv{MM2S}{Memory-mapped to Streaming}
    \abbv{S2MM}{Streaming to Memory-mapped}
    \abbv{FIFO}{First-in-First-out}
    \abbv{NISQ}{Noisy Intermediate-Scale Quantum computing}
    \abbv{VQE}{Variational Quantum Eigensolvers}
    \abbv{QAOA}{Quantum Approximate Optimization Algorithm}
    \abbv{VHDL}{VHSIC Hardware Description Language}
\end{itemize}
}

\clearpage

\section*{Acknowledgment}
This work is supported by a collaboration between the US DOE and other Agencies. This material is based upon work supported by the U.S. Department of Energy, Office of Science, National Quantum Information Science Research Centers, Quantum Systems Accelerator. Additional support is acknowledged from Q-SEnSE, an NSF Quantum Leap Challenge Institute (NSF QLCI Award OMA-2016244).
Sandia National Laboratories is a multimission laboratory managed and operated by National Technology \& Engineering Solutions of Sandia, LLC, a wholly owned subsidiary of Honeywell International Inc., for the U.S. Department of Energy’s National Nuclear Security Administration under contract DE-NA0003525.  This paper describes objective technical results and analysis. Any subjective views or opinions that might be expressed in the paper do not necessarily represent the views of the U.S. Department of Energy or the United States Government.

% This work was funded by the CUbit Quantum Initiative, which includes Q-SEnSE: Quantum Systems through Entangled Science and Engineering (NSF
% QLCI Award OMA-2016244), as well as by the U.S. Department of Energy, Office of Science, National Quantum Information Science Research Centers.
% Sandia National Laboratories is a multimission laboratory managed and operated by National Technology \& Engineering Solutions of Sandia, LLC, a wholly owned subsidiary of Honeywell International Inc., for the U.S. Department of Energy’s National Nuclear Security Administration under contract DE-NA0003525.  This paper describes objective technical results and analysis. Any subjective views or opinions that might be expressed in the paper do not necessarily represent the views of the U.S. Department of Energy or the United States Government.
%\clearpage

%\nocite{*}
% \bibliographystyle{IEEEtran}
%\bibliography{IEEEabrv,bibliography}
\bibliography{bibliography}

%\input{chapters_v2/bio}
%\begin{IEEEbiography}[{\includegraphics[width=1in,height=1.25in,clip,keepaspectratio]{a1.png}}]{First A. Author} (M'76--SM'81--F'87) and 
%\end{IEEEbiography}

%\begin{IEEEbiography}[{\includegraphics[width=1in,height=1.25in,clip,keepaspectratio]{a2.png}}]{Second B. Author} was 
%\end{IEEEbiography}

%\begin{IEEEbiography}[{\includegraphics[width=1in,height=1.25in,clip,keepaspectratio]{a3.png}}]{Third C. Author, Jr.} (M'87) received 
%\end{IEEEbiography}

%\appendix
%\input{chapters/appendix}

% \EOD
\end{document}